\documentclass[12pt,preprint]{aastex}

\shorttitle{Dust Concentration in Inhomogeneous MRI Disk}
\shortauthors{Kato et al.}

\begin{document}

\title{Dust Concentration at the Boundary Between Steady
Super/Sub-Keplerian Flow Created by Inhomogeneous Growth of MRI}

\author{M. T. Kato}
\affil{Department of Earth and Planetary Science, Tokyo Institute of Technology, Ookayama 2-1-12-I2-10, Meguro-ku, Tokyo}
\email{marikok@geo.titech.ac.jp}

\author{M. Fujimoto}
\affil{Institute of Space and Astronomical Science, Japan Aerospace Exploration Agency, Yoshinodai 3-1-1, Sagamihara, Kanagawa}

\and

\author{S. Ida}
\affil{Department of Earth and Planetary Science, Tokyo Institute of Technology, Ookayama 2-1-12-I2-10, Meguro-ku, Tokyo}


\begin{abstract}
How to create planetesimals from tiny dust particles in a proto-planetary disk 
before the dust particles spiral to the central star is one of the most 
challenging problems in the theory of planetary system 
formation. In our previous paper \citep{kato08}, 
we have shown that a steady angular 
velocity profile that consists of both super and sub-Keplerian regions 
is created in the disk through non-uniform excitation of Magneto-Rotational Instability 
(MRI). Such non-uniform MRI excitation is reasonably expected in a part of 
disks with relatively low ionization degree. In this paper, we show through 
three-dimensional resistive MHD simulations with test particles that this radial structure of the angular 
velocity indeed leads to prevention of spiral-in of dust particles and 
furthermore to their accumulation at the boundary of 
super-Keplerian and sub-Keplerian regions.
Treating dust particles as test particles, their motions 
under the influence of the non-uniform MRI through gas drag
are simulated. 
In the most favorable 
cases (meter-size dust particles in the disk region 
with a relatively large fraction of MRI-stable region), 
we found that the dust concentration is peaked around the super/sub-Keplerian 
flow boundary and the peak dust density is 10,000 times as high as the 
initial value. The peak density is high enough for the subsequent 
gravitational instability to set in, suggesting a possible route to 
planetesimal formation via non-uniformly excited MRI in weakly 
ionized regions of a disk.
\end{abstract}

\keywords{protoplanetary disks --- instabilities --- MHD --- 
planetary systems: formation --- turbulence}

\defcitealias{kato08}{Paper~I}


\section{Introduction}

One of the most serious problems in planet formation is how the dust components grow up to larger objects in protoplanetary disks. The known key-stage is when the dust particles are meter-size \citep{adachi76, wei77}. Since gas pressure gradient in a protoplanetary disk is negative, gas rotates at sub-Keplerian velocity. On the other hand, the dust particles tend to rotate at Keplerian velocity and therefore they feel headwind of gas. Losing their angular momenta, they spiral into a central star quickly on timescales of ~100 years.

One possibility to overcome the meter-size barrier is the planetesimal formation via gravitational collapse of a cluster composed with small dust particles \citep{saf69, gold73}. 
The dust would settle to midplane in the absence of turbulence. Early studies of dust settling in turbulent accretion disks were carried out by \citet{cuzzi93}, \citet{miyake95} and \citet{dub95}. They presented that large dust particles settle down toward their equilibrium distribution in a turbulent diffusive time scale while small ones remain mixed throughout the whole gas disk. If dust density in midplane becomes high enough, Kelvin-Helmholtz instabilities between a sediment layer of dust and the overlying gas can drive turbulence and mix the dust layer enough to prevent gravitational instability \citep{wei77, wei80, ishi03, gomez05}. \citet{joh06b}, however, carried out two-dimensional simulation and found active turbulent concentration. Except the effect of Kelvin-Helmholtz instabilities, the high dust density in midplane leads to a streaming instability \citep{good00, you05} because the dust is susceptible to preferential radial migration relative to the gas \citep{naka86}. The streaming instability was shown to lead to turbulence and concentrate dust particles locally \citep{you07, joh07a}. As one of other mechanisms of turbulent concentration, the effect of the turbulence driven by the magnetorotational instability (MRI; \citealp{bal91, haw95}) is recognized. Planetesimal formation by gravitational instability in the MRI turbulence is addressed by \citet{joh06a, joh07b} and \citet{balsara09}. 

Turbulence level is the most fundamental parameter governing whether planetesimals can form by gravitational instability in these models.
The turbulence level of MRI depends on the ionization degree of disk gas and the strength of magnetic field. The linear analysis showed that the growth rate of MRI is reduced by ohmic diffusion when collisions are so frequent that not only the ions and neutrals are well coupled but also electrical currents are damped \citep{jin96, sam99}. At the same time, the growth wavelength becomes longer and transport rate of angular momentum is diminished. The nonlinear evolution of the MRI in nonideal MHD starting from the weak vertical magnetic field has been studied by \citet{sano98}. They found that sustained MHD turbulence requires the magnetic Reynolds number $R_{\rm m}=v^2_{\rm A}/\eta \Omega \gtrsim 1.0$, where $v_{\rm A}$, $\eta$ and $\Omega$ are Alfven velocity, magnetic resistivity and Keplerian rotation frequency.

The values of $\eta$ are regulated by an ionization structure in a protoplanetary disk. The inner regions of a disk are thermally ionized \citep{pne65, ume83}. At greater distances from the disk center, stellar X-rays and diffuse cosmic rays ionize the surface gas layer down to a certain column density \citep{glass97, igea99}. In moderately distant regions in which disk column density is high enough, midplane layers are not highly ionized and MRI could be inactive there. This region is called "dead zone" \citep{gam96}. 

The characteristics and interesting physics are prospective around dead zones. The magnetic turbulence outside the dead zone, which transports angular momentum and mass, creates gas density bump at the outer boundary of the dead zone. This density pile-up triggers the Rossby wave \citep{li01, varni06} to form long-life anticyclonic vortices. \citet{barge95} suggested that the dust particles are trapped in a center of vortex. By numerical simulation of Rossby vortices including solid particles, \citet{ina06} confirmed the enhancement of dust density there. \citet{lyra08} showed that gravitationally bound embryos can form in the regions of enhanced dust density. 

\citet[][; hereafter referred to as Paper I]{kato08} 
showed that non-uniform excitation of MRI creates 
angular velocity profile that consists of super and sub-Keplerian 
regions locally in the disk (for details, see section 2).
It is expected that this radial structure of the angular 
velocity leads to prevention of spiral-in of dust particles and 
their accumulation at the boundary of 
super and sub-Keplerian regions, since 
the super-Keplerian flow pushes the dust particles outward 
while the sub-Keplerian flow drags them inward.
This interesting mechanism is based on the co-existence of MRI active and inactive regions.
The ionization degree that regulates MRI depends on the relative abundance and size distribution of grains \citep{wardle99, sano00}.
It can also be changed by phase chemistry \citep{ilgner08}.
Slight changes in these quantities near boundaries at global active/dead zones of MRI may create the local inhomogenity in MRI growth.  
We will address this issue in a subsequent paper. Note that the local model in this paper is not a toy model that mimics the global inhomogeneous structure of dead and active zones.

In this paper, we investigate dust accumulation induced by inhomogeneous MRI evolution by three-dimensional non-ideal MHD simulation including dust particles. 
The dust particles are represented by Lagrangian particles suffering gas drag and are assumed not to affect on gas. 
We consider both cases with radially non-uniform magnetic resistivity under uniform magnetic field direction and non-uniform magnetic field with radially uniform magnetic resistivity. MRI develops in low resistivity regions in the former case and in strong vertical magnetic field regions in the latter case. The latter model is the same as the model adopted in Paper I. 
We briefly summarize the results of Paper I in section~\ref{sec:paper1}. We explain our simulation setting in section~\ref{sec:model}. In section~\ref{sec:result}, we present that the MRI leads to quasi-steady state and particles are locally concentrated as expected. We also analyze the distributions of particle radial velocity to understand the effect of weak remnant magnetic turbulence. We examine how high the particle density enhancement is and the possibility of planetesimal formation via gravitational instability. Finally, we summarize our results and discuss their implications in section~\ref{sec:discussion}.


\section{Brief Summary of \citetalias{kato08}}\label{sec:paper1}

In Paper I, we have studied how the angular velocity profile of gas is modified when MRI is excited non-uniformly in a part of a disk. Local two-dimensional simulations (in the $x$-$z$ plane where $x$ and $z$ are radial and vertical coordinates to the disk) with the shearing box boundary conditions were performed. We assumed the ionization degree of the gas to be relatively low. 
The vertical component to the disk plane ($z$ component) of magnetic field was changed along the radial direction ($x$) while the absolute values of the field is constant.  In the part of a disk where the ionization degree is low (equivalently, resistivity is high), MRI is active only in the regions with relatively large vertical component of the magnetic field. One of the models used in the present paper has the same setting but with three-dimensional simulations, while the other model in the present paper assumes non-uniform resistivity with uniform magnetic field. 

\begin{figure}
\figurenum{1}
\epsscale{0.6}
\plotone{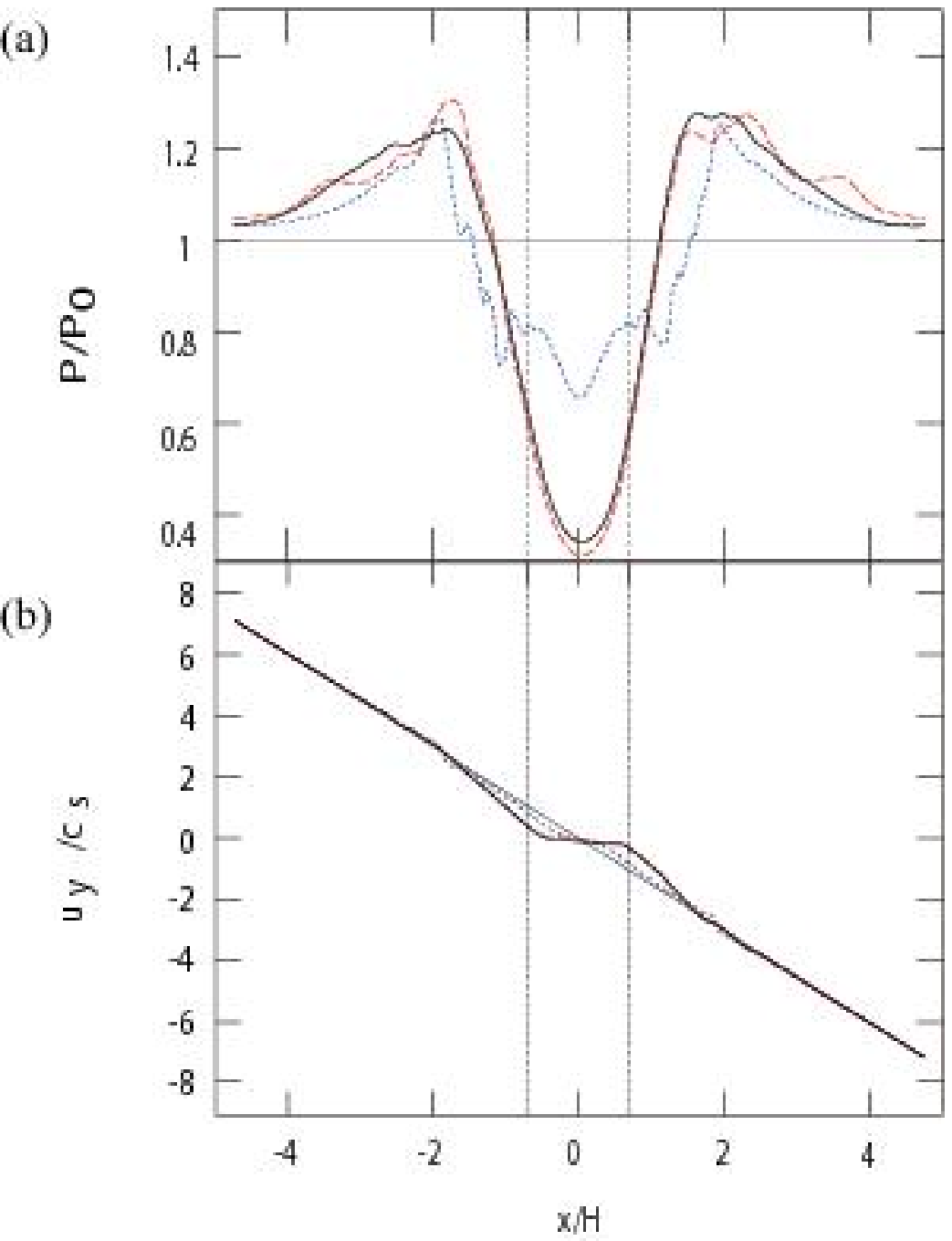}
\caption{Results of the two-dimensional version of model-s40 (${\rm d}B_z(t=0)/{\rm d}x \neq 0, L_{\rm s}/H=4.0$) described in Paper I.
The corresponding result of the three-dimensional simulation is shown in Fig.~7.
 Time evolution of vertically averaged values of (a) pressure $P$ and (b) gas angular velocity $u_{y}$. The thin solid, dotted, dashed and bold lines are the snapshots at $t\Omega=0, 27, 40$ and 70, respectively. The two vertical dotted-lines are the boundaries between the unstable and stable regions. MRI is initially excited only between the two dotted lines. }
\label{fig:paper1}
\end{figure}

The most illustrative result of Paper I is shown in Figure~\ref{fig:paper1}. Time evolution ($t\Omega=0, 40, 70$) of vertically averaged pressure and angular velocity of gas is plotted in Panels a and b. 
MRI is excited only in the initially unstable region ($-0.71<x/H<0.71$) with large enough initial vertical field and vigorous angular momentum and mass transport occurs within the unstable zone. 
When the magnetic perturbations are propagated to the initially stable regions 
($\left|x/H\right|<0.71$), they are dissipated via resistivity and the large part of the stable region remains undisturbed. 
Hereafter, we call the initially MRI stable and unstable regions simply as "stable" and "unstable" regions, respectively.
After the vigorous angular momentum and mass transfer, a new quasi-steady state with the local rigid rotation feature appears in the unstable region (Figure~\ref{fig:paper1}b). 
The pressure profile of the gas is also modulated (Figure~\ref{fig:paper1}a). 
The rigid rotation feature that appears in the saturated quasi-steady state is sustained by the modified pressure pattern with its two peaks located around the unstable-stable boundary regions and its bottom located in the middle of the unstable zone. The quasi-steady rigid rotation pattern results in super-Keplerian angular velocity at the outer part of the unstable region. Since the gas in the stable region is unperturbed and remains sub-Keplerian when the global pressure gradient is taken into account, one finds that super- and sub-Keplerian zones co-exist in the part of the disk due to the non-uniform excitation of MRI. 

The co-existence of super- and sub-Keplerian zones has significant implication for the dust particle dynamics in a disk. As mentioned in the Introduction, the dust particles feel headwind and migrate inward in the sub-Keplerian gas flow. 
On the other hand, the dust particles feel tailwind and migrate outward in the super-Keplerian zone. 
Then at a boundary between sub- and super-Keplerian zones, with the super-Keplerian zone situated on the inner side, which is indeed seen at the outer edge of the unstable region, the inward migration of dust particles through the sub-Keplerian zone (the stable zone) are halted. 
The boundary region potentially acts as a barrier for the dust infall and furthermore as the focal annulus of the dust particle concentration which may eventually lead to planetesimal formation.

The widths of the stable and unstable regions are one of the key parameters that determine the characteristics of the final state. We have found that the final state is classified by the spatially averaged magnetic Reynolds number in the initial condition, 
\begin{eqnarray}
R_{\rm m,ave}=v^2_{{\rm A}z,{\rm ave}}/\eta\Omega, \label{Rmave}
\end{eqnarray}
where $v_{{\rm A}z} = B_z /\sqrt{4 \pi \rho}$ is $z$ component of Alfven velocity.
If $R_{\rm m,ave}$ is less than unity, a quasi-steady state with the local super-Keplerian rotation appears.
The turbulence level within the stable region and at the boundary varies according to $R_{\rm m,ave}$.
The degree of dust concentration at the boundary region would be affected by the level of turbulence. 
For $R_{\rm m,ave}>1$, the magnetic perturbations are not dissipated enough within the stable region and the velocity pattern undergoes perpetual changes without reaching a quasi-steady state.
Steady accumulation of dust particles is not expected in this case. 
Note that our simulations with constant $R_{\rm m,ave}$
converge with increasing resolution, because 
the resistivity $\eta$ (ohmic dissipation) is explicitly included 
in our simulations and it is kept constant due to the constancy of $R_{\rm m,ave}$.

In the present paper, we address the following questions that arise from the results of Paper I: 
(1) Do dust particles truly concentrate at the boundary? 
(2) Is the emergence of the quasi-steady rigid rotation region preserved in three-dimensional cases? (3) How does remaining weak turbulence affect the dust concentration process? (4) Are dust particles concentrated dense enough on long enough timescales for subsequent gravitational instability to set in?


\section{Model}\label{sec:model}


\subsection{Equations}

Here we are concerned with a small region in a protoplanetary disk and study the local dynamics of gas and dust particles. The gas motion is calculated by resistive MHD simulations. The dust particles are modeled as test particles that move under the effects of the gravity from the central star and the gas drag.

We adopt a local shearing box that has radial ($x$), azimuthal ($y$) and vertical ($z$) directions. 
The boundary conditions are periodic in the $y$ and $z$ directions and the shearing box boundary condition is used in the $x$ direction \citep{wisdom88, haw95}.
We calculate compressible resistive magnetohydrodynamic (MHD) equations given by
\begin{eqnarray}
\frac{\partial \mathbf{u}}{\partial t}
+\left(\mathbf{u}\cdot\nabla\right)\mathbf{u}&=&
-\frac{1}{\rho}\nabla\left(P+\frac{\mathbf{B}^2}{8\pi}\right)
+\frac{1}{4\pi\rho}\left(\mathbf{B}\cdot\nabla\right)\mathbf{B}
-2\mathbf{\Omega}\times\mathbf{u}+3\Omega^2x\hat{\mathbf{x}}
-\beta c_{s}\Omega\hat{\mathbf{x}}, 
\label{motion}\\
\frac{\partial \rho}{\partial t}
+\nabla\cdot\left(\rho\mathbf{u}\right)&=&0, \label{continuity}\\
\frac{\partial \mathbf{B}}{\partial t} &=&
\nabla\times\left[\left(\mathbf{u}\times\mathbf{B}\right)-
\eta\left(\nabla\times\mathbf{B}\right) \right], 
\label{induction}\\
P&=&c^2_{s}\rho, \label{isothermal}
\end{eqnarray}
where $\hat{\mathbf{x}}$ is a unit vector in the $x$ direction. 
We consider only the Ohmic dissipation in equation~(\ref{induction}) because influence of ambipolar diffusion is negligible at the density level that is expected in the inner ($\lesssim 100\rm AU$) midplane of a protoplanetary disk \citep{jin96, sam99, chi07}. 
The last terms in r.h.s. of equation~(\ref{motion}) allows the effect of the global pressure gradient to be taken into account in our local model. The parameter $\beta$ is related to the global pressure gradient as
\begin{eqnarray}
-\frac{1}{\rho}\frac{\partial P}{\partial r}
&=&-\frac{1}{\rho} \left(\frac{P}{r} \right) \alpha \\
&=&-\frac{H}{r} \alpha c_{s}\Omega \\
&=&-\beta c_{s} \Omega \left(=-\frac{\beta^2}{\alpha} \Omega^2 r \right), \label{globalP}
\end{eqnarray}
where we assume $P \propto r^{\alpha}$, with constant $\alpha$ and an isothermal disk $c_{s}=H\Omega$, where $H$ is disk scale height. 
The value of $\beta \left(=\alpha H/r =\alpha c_s/r\Omega \right)$ is treated as a constant parameter in our local model.
We solve equation~(\ref{induction}) with MOCCT method \citep{sto92} and the advection terms in equation~(\ref{motion}) with CIP method \citep{yabe91}. 

The dust grains are treated as test particles. 
The equation of motion of the $i$-th particle is given by
\begin{eqnarray}
\frac{{\rm d} \mathbf{v}_{i}}{{\rm d} t}&=&
-2\mathbf{\Omega}\times\mathbf{v}_{i}+3\Omega^2x_{i}\hat{\mathbf{x}}
-\frac{1}{\tau_{f}}\left(\mathbf{v}_{i}-\mathbf{w}_{i}\right), \label{particle}
\end{eqnarray}
where $\mathbf{w}_{i}$ is the gas velocity at the location of the $i$-th particle,
which is interpolated with the three-dimensional first-order interpolation scheme by values of $\mathbf{u}$ at the eight grid points surrounding the particle.

The friction force, the last term in equation~(\ref{particle}), is proportional to the velocity difference between the dust and gas since the dust sizes that we are concerned with are in Epstein and Stokes drag-law regimes.
The characteristic friction (stopping) timescale $\tau_{f}$ is given by
\begin{eqnarray}
\tau_{f}&=&\frac{a^2_{\bullet}\rho_{\bullet}}
{{\rm min}\left(a_{\bullet}c_{s}, 9/2\nu\right)\rho},
\label{friction}
\end{eqnarray}
where $a_{\bullet}$, $\rho_{\bullet}$ and $c_{s}$ are the dust particle radius, dust particle density and the sound speed, respectively. $\nu$ is the molecular viscosity and equal to $c_{s}\lambda/2$, where $\lambda$ is the mean free path of the gas molecules. Depending on the particle size relative to the mean free path ($a_{\bullet} > {\rm or} <\left(9/4\right)\lambda$) the drag law changes from the Epstein to the Stokes regime. The scaled friction times of $\tau_{f}\Omega=0.1$ and $1.0$ correspond to the grain sizes of approximately 0.1 and 1 meter, respectively, at the radial location of Jupiter in the minimum mass solar nebula model.


\subsection{Initial Setup}

All of our simulations start with uniform gas density ($\rho_0$) and gas pressure ($P_0$). The global gas pressure gradient is set to be $\beta=-0.04$. 
The initial gas angular velocity profile is given by unperturbed Keplerian flow ($-(3/2)x\Omega$) and a small reduction due to global gas pressure gradient ($\beta c_{s}/2$), $u_{\rm y}=-(3/2)x\Omega+\beta c_{s}/2$. 
Initial disturbances are given to the gas radial velocity with the amplitude $\left|\delta u_{x}\right|=0.001c_{s}$.
Test particles are initially orbiting at the Kepler angular velocity. The particles move inward in the initial stage because they lose their angular momentum by the headwind that they feel due to the global pressure gradient. This is the dust infall problem described in Introduction. Initially 8 particles per gird are distributed. 
Since typical grid we use is $(250-950) \times 100 \times 50$ (Table~\ref{tab:1}), totally $\sim O(10^7)$ particles are distributed.

The non-uniformity in the MRI growth is set either by non-uniform resistivity (CASE1) or non-uniform vertical ($z$) component of the magnetic field (CASE2).
We describe these cases in details in the following.

\subsubsection{CASE1}

In CASE1, gas ionization degree is set to be radially inhomogeneous under the uniform magnetic field. 
The linear analysis for MRI caused by vertical magnetic field shows that the larger resistivity makes unstable wavelength longer \citep{jin96, sam99}. 
The upper limit to the wavelength, $\lambda_{z, {\rm crit}}$, is scale height for actual disks while in our simulation, it is the size of the simulation box in the $z$ direction ($L_{z}$). The critical value of resistivity with which the modes with wavelength shorter than $\lambda_{z, {\rm crit}}$ cannot grow is
\begin{equation}
\eta_{\rm crit} \simeq \frac{1}{k_{z, {\rm crit}}} \sqrt{ \frac{2}{\beta_{plasma}} \left( \frac{3\Omega^2}{c_{s}^2}-\frac{2k_{z, {\rm crit}}^2}{\beta_{plasma}} \right)}H^{2}\Omega,
\label{eq:eta.crit}
\end{equation}
where $k_{z, {\rm crit}}$ is the wave number corresponding $\lambda_{z, {\rm crit}}$. Here $\beta_{plasma} ={2c^2_{s}}/v^2_{{\rm A}z}$ is the plasma beta. In our case there is no growing mode if the resistivity makes $\lambda_{z,{\rm crit}}$ larger than $L_{z}$. 

\begin{figure}
\figurenum{2}
\epsscale{0.6}
\plotone{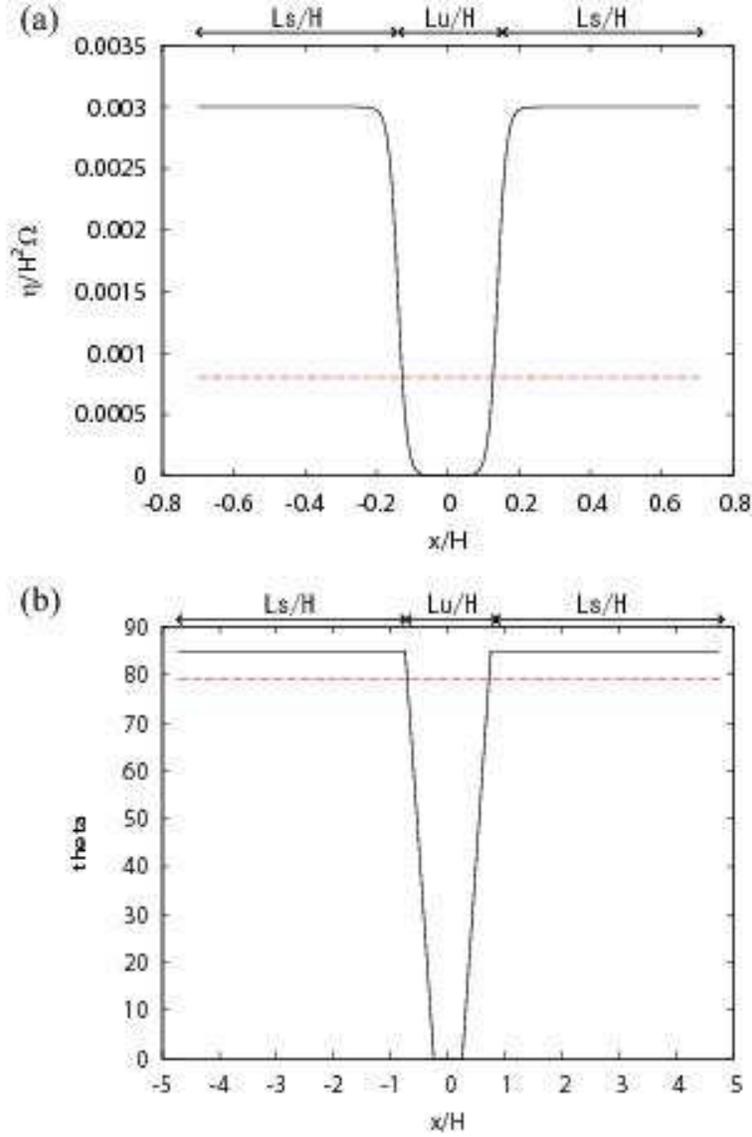}
\caption{Two versions of initial setting to realize the non-uniform MRI growth in 
the simulation box. (a) The non-uniform radial distribution of resistivity used in CASE1 (model-$\eta$), where the dashed line represents the critical value for MRI development predicted by the linear theory. (b) The non-uniform radial distribution of the magnetic field angle $\theta$ in CASE2 (model-s40), where the dashed line shows the critical angle for MRI predicted by the linear theory (for detail, see Paper I). }
\label{fig:ini}
\end{figure}

In CASE1, we set $\beta_{plasma}=4000$ and $\mathbf{B}_0=(0, 0, B_{z})$.
Resistivity varies in the radial direction such that MRI grows only in a limited zone in the center of the simulation box. 
The resistivity distribution is not changed throughout runs. 
The radial distribution is shown in Figure~\ref{fig:ini}a. 
In this paper, the zone at the center where resistivity is initially sub-critical is referred to as the "unstable" region. The radial width of the stable and unstable regions are denoted by $L_{\rm s}$ and $L_{\rm u}$, respectively. 

We have performed only one simulation run for CASE1, which will be called model-$\eta$ in the later sections. Table~\ref{tab:1} lists the parameter values used in the run.

\subsubsection{CASE2}

The set-up of CASE2 is the same as that in Paper I except for three-dimensional simulations. The magnetic resistivity is uniform but the vertical component of magnetic field varies in the radial direction. That is, the initial magnetic field is situated as $\mathbf{B}_0=(0, B_{0}\sin\theta, B_{0}\cos\theta)$, where $\theta=\theta(x)$ is the angle between the vertical axis and the magnetic field. The vertical magnetic component, or $\theta$, determines whether the instability grows or not. 
For large $\theta$, vertical magnetic field is weak and short vertical wavelength modes grow faster. Since such modes are dissipated by the ohmic dissipation effectively, the vertical field MRI does not develop. 
The MRI caused by the azimuthal magnetic field is known to grow rapidly with the limit of extremely short vertical wavelength in ideal MHD \citep{bal98} but the such short vertical wavelength modes tend to be damped resistively. In fact, \citet{papa97} found that the azimuthal field MRI growth ceases for the ionization degree higher than that stabilizes the vertical field MRI. We also did not find the development of the azimuthal field MRI in our simulation as long as the box size in the $y$ direction is expanded by a factor up to 3 from the nominal cases at least for about 20 orbits in which the gas velocity and dust density vary dramatically.

The radial distribution of $\theta$ is shown in Figure~\ref{fig:ini}b. Here, $\beta_{plasma}=400$, and the magnetic resistivity is $\eta=0.002H^2\Omega$. The larger vertical magnetic component in the center of the box makes MRI locally unstable. Outside the unstable zone the value of $\theta$ is smaller than the critical one and MRI is stabilized. 

The two-dimensional version of CASE2 was studied in Paper I, in which 
we found that if the spatially averaged magnetic Reynolds number $R_{\rm m,ave}$ (Equation~(\ref{Rmave})) is small enough, the quasi-steady angular velocity profile that is deviated significantly from the Keplerian is generated. 
Since it is expected that this feature is preserved in three dimensional simulations and $R_{\rm m,ave}$ depends on the radial widths of the unstable region $L_{\rm u}$ and the stable region $L_{\rm s}$, we investigate several cases with various $L_{\rm s}$ while keeping $L_{\rm u} = 1.4H$ in CASE2: $L_{\rm s}=4.0H$($R_{\rm m,ave}=0.096$, model-s40), $1.1H$($R_{\rm m,ave}=0.37$, model-s11), $0.55H$($R_{\rm m,ave}=0.64$, model-s055). 
Note that two-dimensional calculations with the same $L_{\rm s}$ and $L_{\rm u}$ were performed in Paper I. 

In all of these cases, the friction time is set to be $1.0/\Omega$. To investigate the effects of the friction time, we set it to be $0.1/\Omega$ in model-t01, which has $L_{\rm u}=1.4H$ and $L_{\rm s}=4.0H$ (the same as model-s40). The initial settings are summarized in Table~\ref{tab:1}.


\subsection{Estimation of Particle Drift Speed}

When the condition $R_{\rm m,ave}<1$ is met, the non-uniform growth of MRI will create a new quasi-steady gas angular flow pattern that should affect the dust particle dynamics substantially (Paper I). 
At the same time, weak remnant turbulent flow may be superposed upon the 
quasi-steady pattern and would affect the dust motions, with its degree being dependent on the spatially averaged magnetic Reynolds number. Here we evaluate these two effects on the dust accumulation. We will be focusing on the $x$ component of the dust velocities $v_x$.

If the turbulence is small enough, as we have mentioned in section~\ref{sec:paper1}, the velocity difference in the $y$ component (angular velocity) between dust and gas is the determining factor for the assemblage of dust particles at the boundary between super and sub-Keplerian zones. 
In order to quantitatively analyze the effect of the remnant turbulence, we divide the radial component of the particle velocity into two components: one is that due to the non-Keplerian gas motion in the quasi-steady state achieved at the MRI saturation and the other is that due to turbulence.
The equation of motion for the dust velocity field at the grid point is
\begin{eqnarray}
\frac{{\rm d} \mathbf{v}}{{\rm d} t}&=&
-2\mathbf{\Omega}\times\mathbf{v}+3\Omega^2x\hat{\mathbf{x}}
-\frac{1}{\tau_{f}}\left(\mathbf{v}-\mathbf{u}\right). \label{particle2}
\end{eqnarray}
We set $\mathbf{v}=\mathbf{v'}+v_{\rm kep}=\mathbf{v'}-(3/2)x\Omega\hat{\mathbf{y}}$ where $\mathbf{v'}$ is a deviation from Kepler flow.
Assuming $d\mathbf{v'}/dt = 0$ ($d\mathbf{v}/dt=-(3/2)\Omega\hat{\mathbf{y}}$) in $x$ and $y$ components and eliminating $v'_y$, the $x$ component of the particle velocity is given by
\begin{eqnarray}
v_{x} \sim v_{\rm d}&=&\frac{2\delta u_{y}}
{\tau_f\Omega+\left(\tau_f\Omega\right)^{-1}}
+\frac{u_{x}}{\left(\tau_f\Omega\right)^{2}+1} \nonumber \\
&=&v_{\rm f}+v_{\rm t}, \label{eq:turb}
\end{eqnarray}
where $\delta u_{y}$ is the gas angular velocity relative to Keplerian velocity, $\delta u_y =u_{y}-v_{\rm kep}$ and $u_{x}$ is the gas radial velocity due to turbulent viscosity. 
The component $v_{\rm f}$ represents the particle velocity
due to non-Keplerian gas motion in the quasi-steady state,
because it does not vanish even in the non-turbulence case, $u_{x}=0$. 
Although turbulence also contributes to $v_{\rm f}$,
the contribution due to turbulence can be neglected compared with
that due to the quasi-steady non-Keplerian gas flow,
except for strong turbulence case where the quasi-steady flow is not achieved.
As we will show below, the component $v_{\rm t}$ can be regarded as
characteristic amplitude of the velocity dragged by turbulence,
although $v_{\rm t}$ changes with time due to the temporal
change of $v_x$ and the assumption of $d\mathbf{v'}/dt = 0$ 
is no more valid in the turbulent case.

Later in this paper, $v_{\rm f}$ and $v_{\rm t}$ will be calculated by 
equation~(\ref{eq:turb}), using gas velocity obtained by
the MHD simulation, and they will be presented as a function of $x$
after azimuthal and vertical averaging. 
The ratio of the absolute values of the two is
\begin{eqnarray}
\frac{\left|v_{\rm t}\right|}{\left|v_{\rm f}\right|}
&=&\frac{\left|u_{x}\right|}{2\left|\delta u_{y}\right|}
\frac{1}{\tau_f\Omega}.\label{eq:ratio}
\end{eqnarray}
The turbulence effect is indicated by this ratio.
It is stronger, 
if i) turbulent flow is larger than the angular velocity difference $\delta u_{y}$, and/or ii) the dust size ($ \tau_f $) is smaller. 
In the annulus where dust particles are expected to accumulate,
$\delta u_{y} \simeq 0$ and the ratio diverges.
However, we will show below that this divergence is restricted
only in narrow regions and the effect of turbulence 
does not prevent dust accumulation.

We have set the global gas pressure gradient to be $\beta=-0.04$ so that a direct comparison with the result by \citet{joh07b} could be performed. This global pressure gradient, however, is less steep than in the Minimum Mass Solar Nebular model \citep{hayashi81}. 
A steeper global pressure gradient ($\beta <-0.04$) has at least three effects in the current context: 
(1) The baseline of gas velocity becomes lower than the Keplerian velocity and it becomes more difficult to create super-Keplerian regions whose outer-edges accumulate the dust particles.
(2) If a super-Keplerian region appears, accumulation towards its outer-edge is faster ($v_{\rm f}$ is faster) for steeper pressure gradient. 
This has a positive impact on the dust accumulation efficiency. 
(3) The outer-edge of super-Keplerian region would change its location according to the value of $\beta$. 
The turbulence level at the focal annulus of dust accumulation varies accordingly. If the turbulence level is higher, it by itself has a negative impact. In CASE2, we have also carried out the runs with $\beta=-0.10$ and found that the trapping of dust particles does occur and its efficiency is slightly higher.


\section{Result}\label{sec:result}

\subsection{Result of CASE1}

In the model-$\eta$, the non-uniform excitation of MRI is realized by non-uniform resistivity while the magnetic field is set uniform. 
The result of model-$\eta$ nicely illustrates our scenario for dust accumulation.

\begin{figure}
\figurenum{3}
\epsscale{1.0}
\plotone{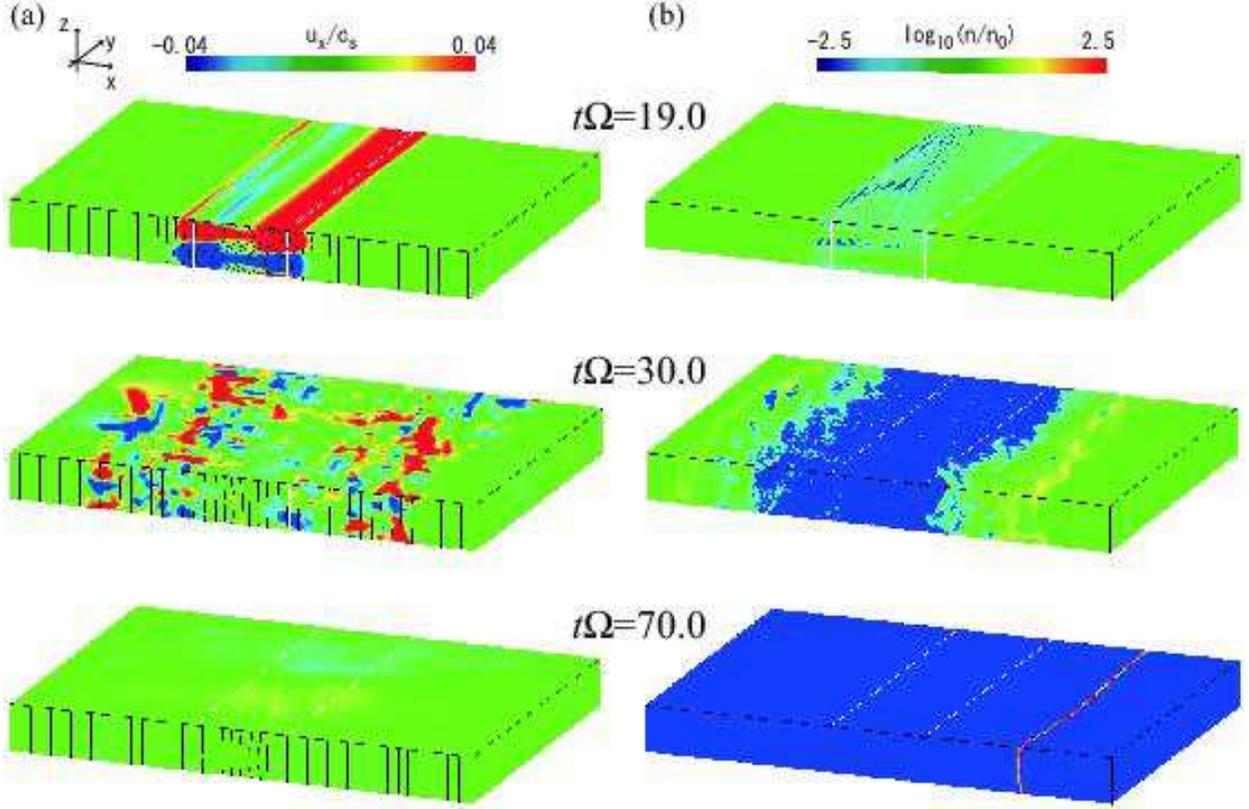}
\caption{The results of model-$\eta$ (${\rm d}\eta/{\rm d}x \neq 0, \tau_f\Omega=1.0$). (a)Time evolution of the magnetic field (black lines) and the angular velocity $u_y$ (contours) of gas.  (b) Particle density normalized by the initial value. The unstable region is between the two white lines. Non-uniform growth of MRI and its relaxation seen in Panel a lead to the dust concentration depicted in Panel b.}
\label{fig:eta-3D}
\end{figure}

Figure~\ref{fig:eta-3D}a shows the evolution of MRI. The black lines depict the magnetic field lines and the gray scale shows the gas radial velocity. The unstable region lies between two white lines ($\left|x/H\right|=0.18$). 
MRI is first excited only in the initially unstable region (see the plot at $t\Omega=19$) and significant angular momentum and mass are transported there.
The MRI turbulence intrudes into the stable region ($t\Omega=30$). Deep inside the stable region, however, is always undisturbed ($\left|x/H\right| \gtrsim 0.5$) because of the rapid dissipation by the enhanced resistivity. 

\begin{figure}
\figurenum{4}
\epsscale{1.0}
\plotone{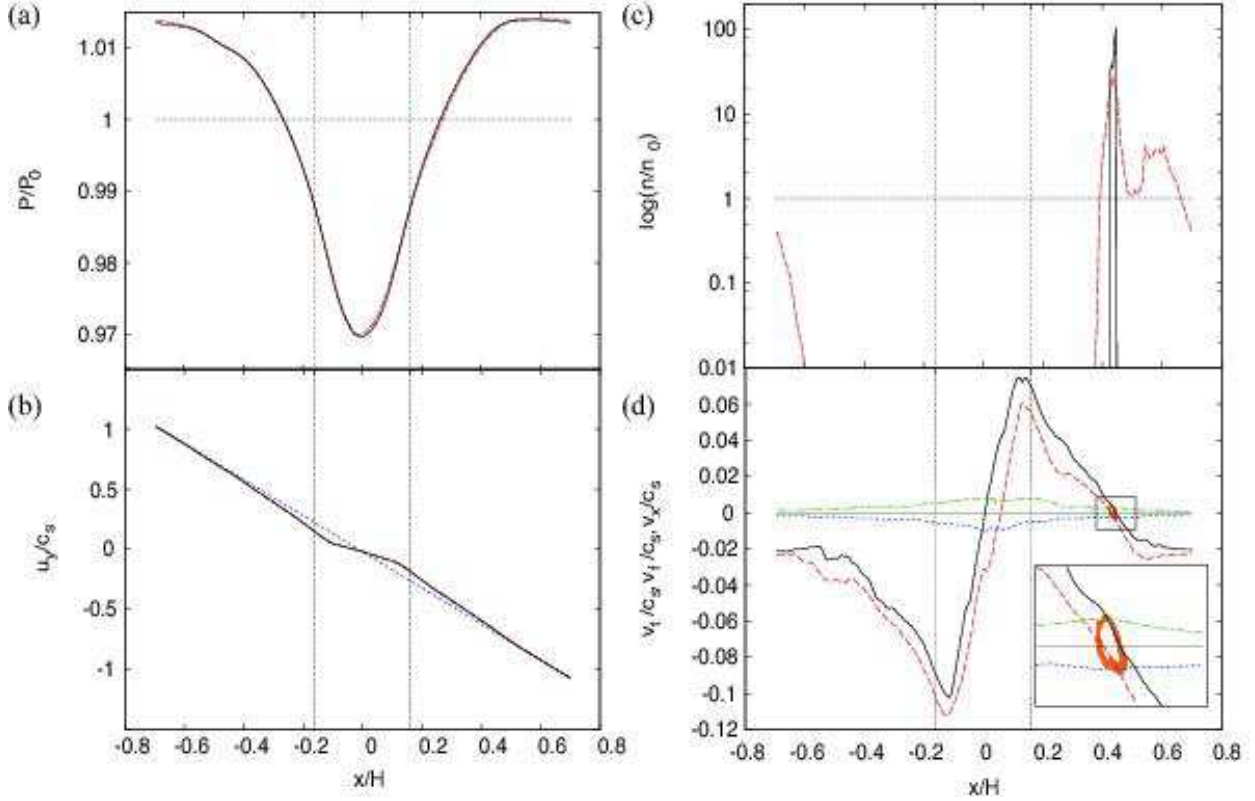}
\caption{Radial dependences of (a) pressure $P$, 
(b) gas angular velocity $u_y$,
(c) the number of particles normalized by the initial value,
and (d) the maximum and minimum values of $v_{\rm f}$ and
$v_{\rm t}$ and the actual particle radial velocity $v_{x,i}$,
in model-$\eta$ (${\rm d}\eta/{\rm d}x \neq 0, \tau_f\Omega=1.0$). 
They are averaged both azimuthally and vertically. 
In Panels a, b and c, the dotted, dashed, and solid lines represent the results at $t\Omega= 0, 40$ and $70$, respectively.
Since $v_{\rm f}$ is predicted
by equation~(\ref{eq:turb}) using simulated gas velocity
after the establishment of the quasi-steady flow ($t \Omega = 70$)
and the simulated gas velocity has a component of turbulence,
it has the maximum (the solid line) and minimum (the dashed line)
values as shown in Panel d.
In Panel d, the maximum/minimum values of 
$v_{\rm t}$ and the actual particle velocity 
are also plotted by dash-dotted/dotted lines and dots.
}
\label{fig:eta-puy}
\end{figure}

Figures~\ref{fig:eta-puy}a and \ref{fig:eta-puy}b show the radial profiles of the pressure and the angular velocity of the gas, respectively. The quantities have been averaged azimuthally and vertically. The sampling times are $t\Omega=$ 0, 40, and 70. 
The zone between the two vertical dotted lines is the unstable region. The inhomogeneous MRI growth creates the rigid rotation pattern in the middle of the simulation box ($\left|x/H\right| \lesssim 0.15$). The pressure distribution is considerably modified such that the resultant pressure gradient force balances with the modified Coriolis force. 
The flattened rotation profile cannot sustain the excitation of MRI in the unstable zone. Indeed, the turbulence weakens extremely at $t\Omega=70$ in Figure~\ref{fig:eta-3D}a. 
The profiles of pressure or angular velocity in Figures~\ref{fig:eta-puy}a and \ref{fig:eta-puy}b depict the quasi-steady state set up by the non-uniform MRI activity. Most of what we see here is quite similar to the results of the two-dimensional simulations described in Paper I even though the initial settings for seed magnetic field and resistivity distribution are totally different.

The rigid rotation causes gas to rotate faster than Keplerian velocity in $0.0\lesssim x/H\lesssim 0.4$ (Figure~\ref{fig:eta-puy}b). This can change the particle migration drastically. Figure~\ref{fig:eta-3D}b shows the temporal evolution of the particle density. The color code is set such that the maximum is ten times the initial value. 
After the particles are swept out of the unstable region by the MRI flow, 
they accumulate to the location at $x/H\simeq0.4$ (note that particles leaving the simulation box from the left hand boundary reenter from the right hand boundary after the shearing box correction is taken into account). 
Though not visible in the panels, the particles initially in the stable zone are swept likewise towards the same location.
The accumulation of particles is most clearly shown in Figure~\ref{fig:eta-puy}c in which the radial distribution of the number of particles that is averaged azimuthally and vertically and is normalized by the initial value. 

To analyze the particle concentration dynamics in more details, 
in Figure~\ref{fig:eta-puy}d, we plot the maximum (the solid line) and 
the minimum (the dashed line) values of $v_{\rm f}$ at a given $x$.
The radial velocities of particles are predicted
by equation~(\ref{eq:turb}) using simulated gas velocity
after the establishment of the quasi-steady flow ($t \Omega = 70$).
Since the gas velocity is influenced by remnant turbulence,
which has variations dependent on $y$ and $z$,
the predicted radial velocities have the maximums and the minimums.
Figure~\ref{fig:eta-puy}d shows that the turbulent fluctuations 
are sufficiently small compared with the systematic angular velocity
change due to the modulation of the pressure profile.
In Figure~\ref{fig:eta-puy}d, the max/min of radial amplitude of turbulent velocity, $v_{\rm t}$, is plotted. 
The amount of $v_{\rm t}$ is smaller than those of $v_{\rm f}$ 
except in the region $\left| v_{\rm f} \right| \approx 0$.
Thus, $v_{\rm f}$ represents the typical radial velocity of particles that is induced by head/tail angular wind in the absence of turbulence. 
Also plotted by dots are the actual particle radial velocities in the simulation result, $v_{x}$,
which are concentrated near the super/sub-Keplerian boundary.
The data are taken at $t\Omega=70$, but it stays essentially the same after the MRI saturation at $t\Omega\simeq40$. 
In the radial regions where the maximum $v_{\rm f}$ is negative, gas rotation is sub-Keplerian for all $y$ and $z$.
In such radial locations, all the particles lose their angular momentum by the drag due to the slower angular wind and migrate inward. 
Conversely, in the regions where the minimum $v_{\rm f}$ is positive, all the particles migrate outward. 
In the middle of the stable zone ($\left|x/H\right| \gtrsim 0.5$), 
the gas flow is hardly changed.
The value of $v_{\rm f} \sim -0.02c_{s}$ corresponds to
the infall speed due to global pressure gradient 
under the initial condition.

The particles are concentrated to the zone where both min[$v_{\rm f}] < 0$ and max[$v_{\rm f}] > 0$ are satisfied, which is situated around $x/H \sim 0.4$. The zone satisfying the two inequalities is rather narrow in $x$, and the difference between the maximum and the minimum $v_{\rm f}$ in this zone is small, resulting in high concentration of the dust particles at the outer-edge of the super-Keplerian zone as shown in Figure~\ref{fig:eta-puy}c. 
Because $\left| v_{\rm t}/(v_{\rm f} + v_{\rm t}) \right| \ll 1$ 
in most of the regions except for the dust concentration zone,
and the width of the dust concentration zone is small, 
the effects of turbulence do not significantly expand the dust concentration zone. 

\begin{figure}
\figurenum{5}
\epsscale{0.6}
\plotone{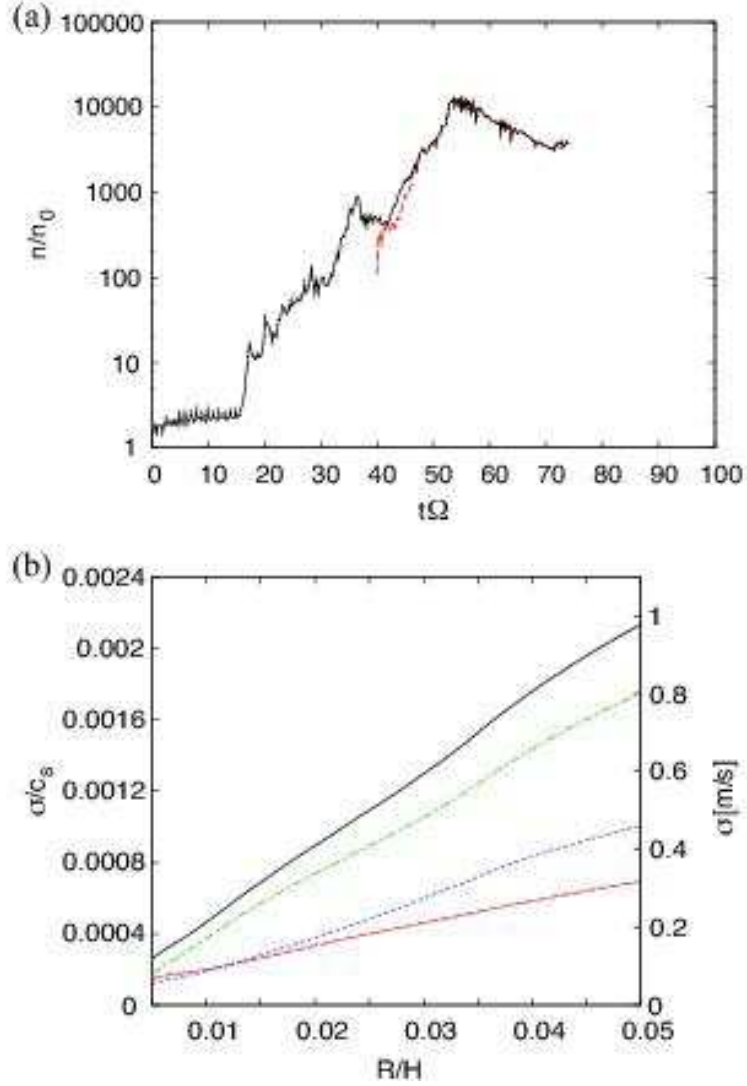}
\caption{Time evolution of dust concentration in model-$\eta$ (${\rm d}\eta/{\rm d}x \neq 0, \tau_f\Omega=1.0$). (a) The solid line represents the number of particles in the cell having the highest density in the whole region, which is normalized by the initial value, while the dashed line is the number of particles in the cell at the center of the traced clump. (b) Velocity dispersion in the traced clump at $t\Omega=t_{c}\Omega$ ($=$55.0 in this case). The solid, dashed, dash-dotted and dotted lines show total, radial, azimuthal and vertical components, respectively. The sound velocity is estimated as 500 m/s with temperature $T\simeq70K$ at 5AU. }
\label{fig:eta-clump}
\end{figure}

In Figure~\ref{fig:eta-clump}a, we plot the time variation of the maximum number of particles per grid. The more or less monotonic increase leads to the peak density of more than 1,000 times the initial density for $t\Omega \gtrsim 50$. 
We will show that the same clump grows in density while keeping its identity, which is needed for subsequent gravitational instability. 

We pick up a particular time $t_c$ during the period when the number of dust particles in the densest grid in the whole region increase monotonically or is saturated.  Then, we search for the cell which has the largest number of particles in it. This cell is identified as the center of the most prominent clump that is composed of a number of cells. The ID numbers of the $N_c$ particles in this cell of the highest density at $t = t_c$ are recorded and their motions are traced in time (also backward in time if necessary). To determine a new position of the center of the same clump at different times, we inspect the cells within 5 grid distance ($0.05H$) from the center of mass of the traced particles. The cell having the largest number of particles among them is defined to be the new position of the center of the clump. Table~\ref{tab:2} shows $t_{c}$ and $N_{c}$ for the runs described in this paper.
The time evolution of the number of particles in the cell at the center of the traced clump is plotted by the dotted lines in Figure~\ref{fig:eta-clump}a. It agrees with the maximum number of particles over the whole region, indicating that the same clump keeps its identity and stays to be the most prominent clump. 

Another way to confirm that the clump steadily holds together is to inspect the velocity dispersion of particles.  Figure~\ref{fig:eta-clump}b presents the velocity dispersion of particles as a function of distance from the center of the clump at $t=t_c$. All the compositional velocity dispersions are extremely low ($\lesssim 0.5 \rm{m/s}$), indicating that the particles are not to be diffused significantly. When the self-gravity is taken into account, the high mass density and the low velocity dispersion should be the preferable condition for the subsequent planetesimal formation via gravitational instability. 

When the global pressure gradient is steeper, the effects on dust dynamics described at the final part of Section 3 must be considered. For the case shown here, we make discussion by inspecting Figure~\ref{fig:eta-puy}d and equation~(\ref{eq:turb}). 
From the maximum amplitude of min[$v_{\rm f}$] change ($\sim 0.08c_s$), we can expect that the value of min[$v_{\rm f}$] remains positive (vertically and azimuthally uniform super-Keplerian region survives) if $\beta>-0.16$. That is, dust particles are prevented from infall as long as $\beta>-0.16$. This condition would be marginally satisfied at $r \la 5$AU in the MMSN model. 
Though the dust concentration that occurs at $v_{\rm f} \simeq0$ would be shifted inward with the more downward shift of the $v_{\rm f}$ (larger negative $\beta$), the turbulence effect on dust concentration would not change because the turbulence level $v_{\rm t}$ is evenly low. 
Therefore, with a steeper global pressure within the range of $\beta>-0.16$, the dust particles would drift faster towards and accumulate more efficiently at the outer-edge of super-Keplerian region.


\subsection{Result of CASE2}

In the models of CASE2, the non-uniform excitation of MRI is realized by non-uniform distribution of the magnetic field while resistivity is set uniform. In the presence of the substantial dissipation, the variation of the vertical component of the magnetic field switches the MRI condition from stable to unstable. In CASE2, the magnetic field angle is varied along the $x$ (radial) direction such that MRI unstable region is situated only in the middle of the simulation box. The overall MHD dynamics varies according to the spatially averaged magnetic Reynolds number ($R_{\rm m,ave}$) which is controlled by the width of the unstable/stable regions in the simulation box, as shown in Paper I. 
Here, the rotation pattern of the magnetic field and the width of the unstable region ($L_{\rm u} =1.4H$) is fixed. 

\subsubsection{model-s40}

\begin{figure}
\figurenum{6}
\epsscale{0.9}
\plotone{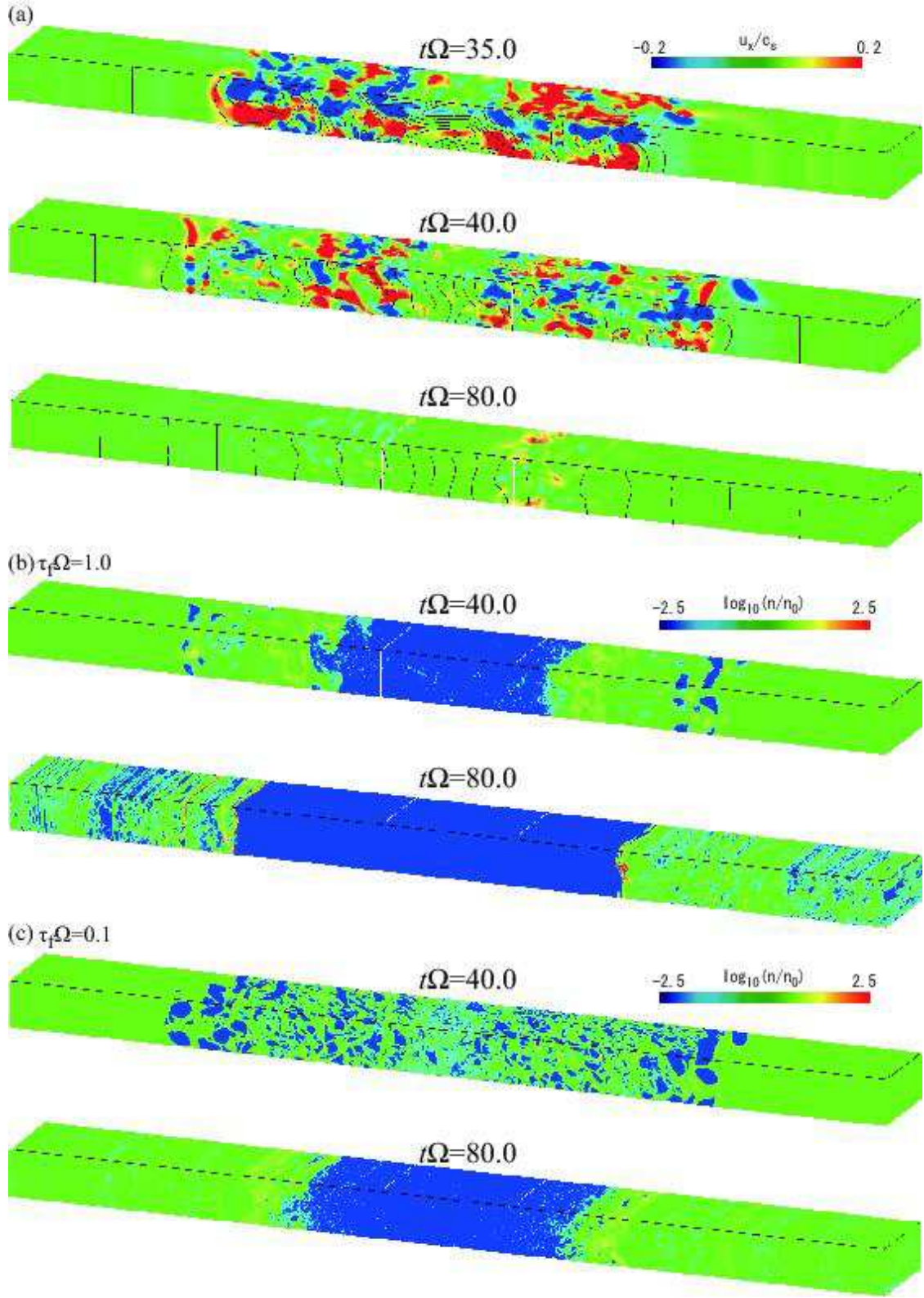}
\caption{ (a)(b) The same as Figure~\ref{fig:eta-3D} but for model-s40 (${\rm d}B_z(t=0)/{\rm d}x \neq 0, L_{\rm s}/H=4.0, \tau_f\Omega=1.0$).
(c) Particle density distribution for model-t01 ($\tau_f\Omega=0.1$).}
\label{fig:s40-3D}
\end{figure}

In the model-s40, the radial width of initial stable region $L_{\rm s}$ is $4.0H$.
The relatively wide stable region results in 
$R_{\rm m,ave}=0.096 (\ll 1)$, which ensures effective enough
dissipation to realize the quasi-steady state. 
Figure~\ref{fig:s40-3D}a presents the evolution of MRI in model-s40. The gray scale and lines are radial velocity of gas and magnetic field lines, respectively. The boundaries between the stable and unstable regions are depicted by the two white lines ($\left|x/H\right|\simeq 0.7$). It is found that the results of the present three-dimensional run are similar to the corresponding two-dimensional results described in Paper I.
The instability is excited only inside the unstable region.
A part of the stable area is disturbed as well, because the resistivity is not large enough to dissipate the fluctuations immediately ($t\Omega=35$). They are, however, eventually dissipated ($t\Omega=40$ and $80$). 

\begin{figure}
\figurenum{7}
\epsscale{0.6}
\plotone{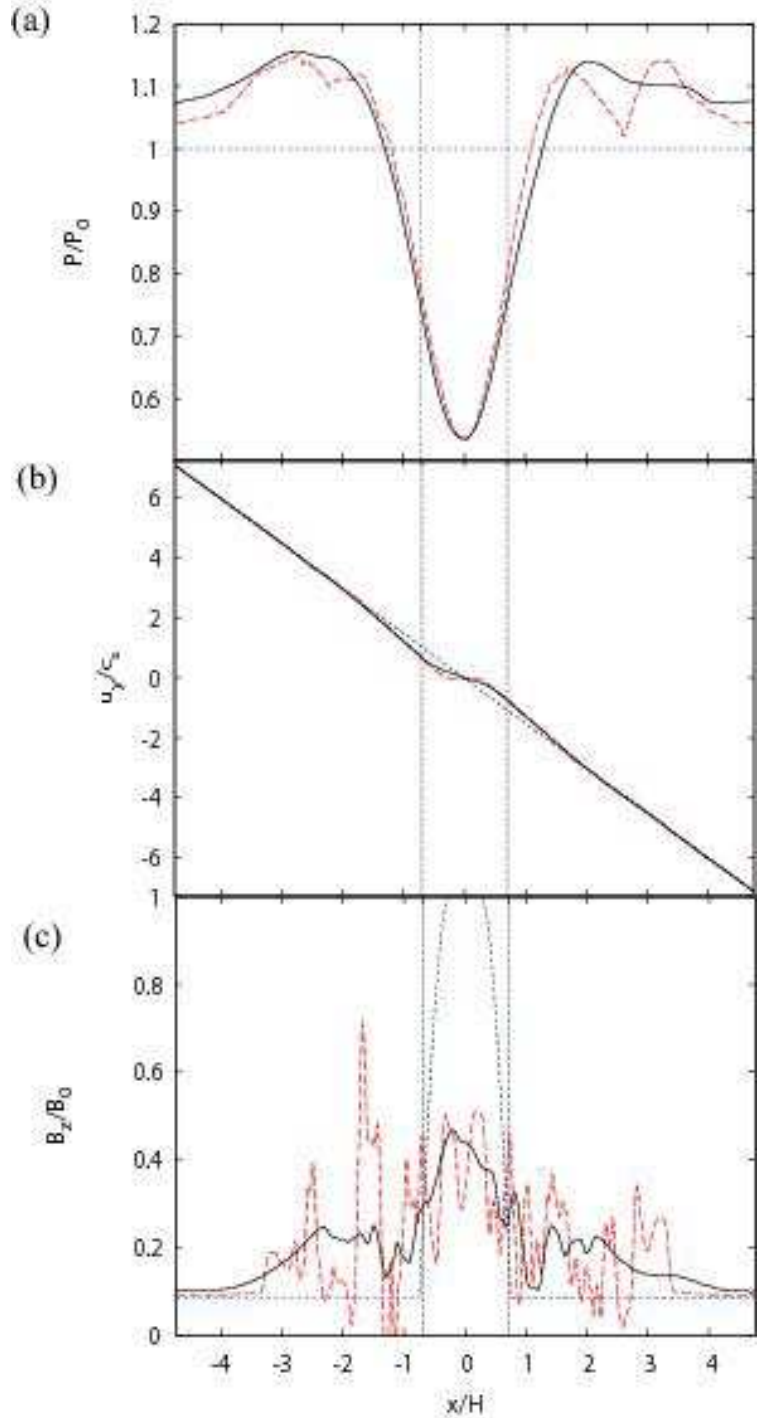}
\caption{Time evolution of azimuthally and vertically averaged values of 
(a) pressure $P$, (b) gas angular velocity $u_{y}$, and (c) vertical magnetic component $B_{z}$, for model-s40 and model-t01 ($L_{\rm s}/H=4.0$). 
The dotted, dashed and bold lines represent the snapshots at $t\Omega= 0, 40$ and $70$, respectively.}
\label{fig:s40-puy}
\end{figure}

Figure~\ref{fig:s40-puy} presents radial distributions of
pressure and angular velocity of gas, and the vertical component of the magnetic field. All the quantities obtained at $t\Omega=$ 0, 40 and 70, are averaged both azimuthally and vertically. The inhomogeneous MRI growth creates the near-rigid rotation pattern in the middle ($\left|x/H\right|\lesssim0.6$) and the modified Coriolis force balances with the newly created pressure gradient force. The small vertical magnetic field together with the flat angular velocity profile suppresses the growth of MRI at later times inside the unstable region in which
MRI was initially excited.

While these results of model-s40 are similar to those of model-$\eta$ (CASE1), three differences are pointed out: 
i) The pressure and angular velocity are distributed less smoothly in the stable region in Figures~\ref{fig:s40-puy}a and \ref{fig:s40-puy}b, which is attributed to deeper penetration of the MRI activity, 
ii) super-Keplerian region exists at $x/H \simeq -2.5$, in addition to
the middle region of $0.0 \lesssim x/H \lesssim 2.0$, 
and iii) the weak instability persists at least until $t\Omega=90$ in the stable region near the unstable region (Figure~\ref{fig:s40-3D}a). 
The super-Keplerian region at $x/H \simeq -2.5$ arises,
because larger stable region makes mass and angular momentum transport 
less efficient around the middle of the stable region ($x/H \sim \pm 4.5$).
As a result, the pressure gradient becomes positive at $x/H \simeq -2.5$.
In model-$\eta$, the large magnetic resistivity in the stable region 
suppresses the instability even in the region of steep radial gradient
of angular velocity.
But, since in the models of CASE2, constant magnetic resistivity 
is assumed, the magnetic field is not completely dissipated
in the stable region with the steep radial gradient.

\begin{figure}
\figurenum{8}
\epsscale{0.6}
\plotone{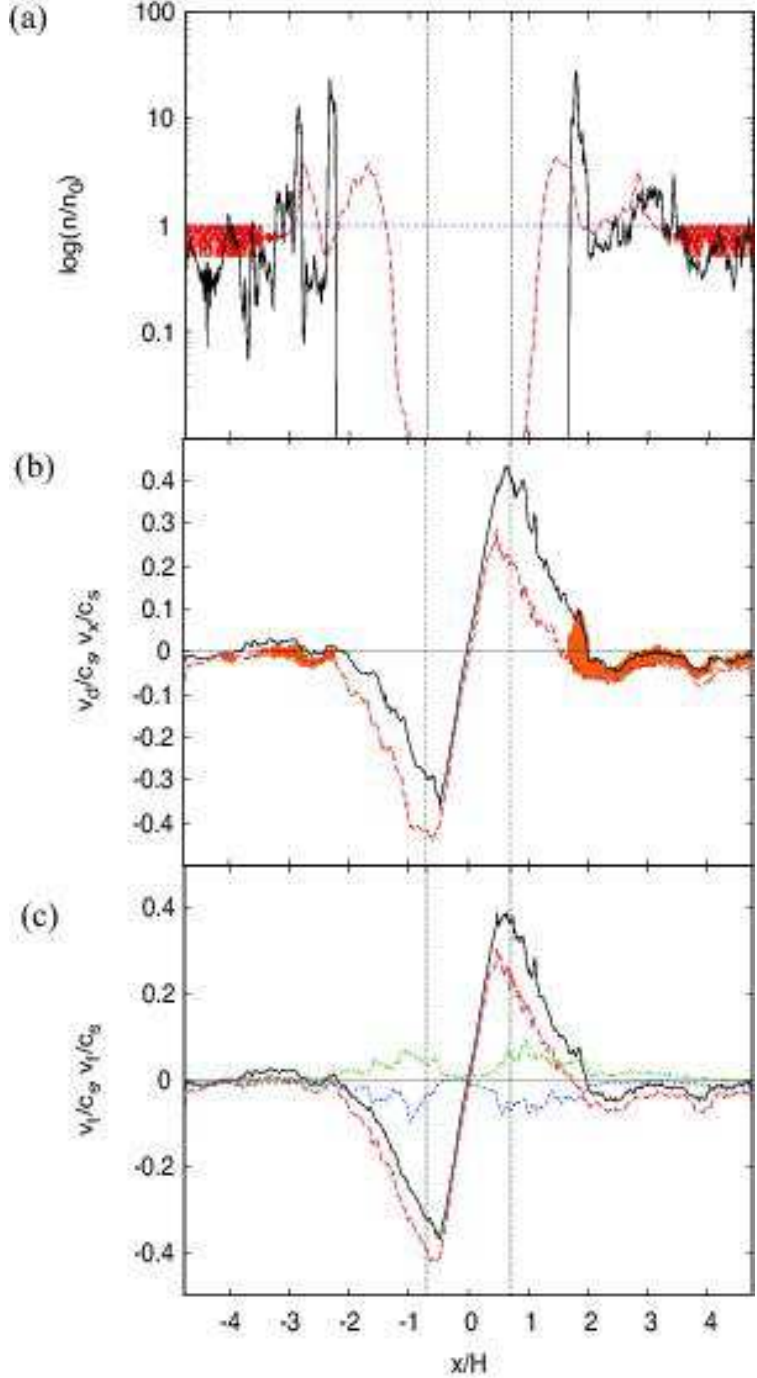}
\caption{Dust concentration in model-s40 ($L_{\rm s}/H=4.0, \tau_f\Omega=1.0$).
(a) Time evolution of the number of particles in each radial grid scaled by the initial value, where the dotted, dashed and bold lines represent the snapshots at $t\Omega= 0, 40$ and $70$, respectively. 
(b) Radial velocity of particle $v_{x}$ (dots) and the estimated radial velocity $v_{\rm d}$ that are azimuthally and vertically averaged, where the solid and dashed lines express the maximum and minimum values in the averaging, respectively. (c) $v_{\rm f}$ (max: bold solid, min: bold dashed) and max/min of $v_{\rm t}$ (thin lines). The results in Panels b and c are obtained at $t\Omega=70$. }
\label{fig:s40-dust}
\end{figure}

Figure~\ref{fig:s40-3D}b presents the temporal evolution of the particle density. The evolution of the particle density pattern is similar to CASE1. However, unlike CASE1, not all the particles are assembled to the focal annulus. 
Figure~\ref{fig:s40-dust}a shows the azimuthally and vertically averaged value of the number of particles as a function of $x$. As expected, particles are condensed at $x/H \simeq 2.0$ (the outer-edge of the super-Keplerian region). Figure~\ref{fig:s40-dust}a also shows some other peaks in the stable region,
which reflects the characteristics i) and ii) of the flow pattern described above. 

Figure~\ref{fig:s40-dust}b shows radial velocities of particles predicted
by equation~(\ref{eq:turb}) using simulated gas velocity after the establishment
of the quasi-steady flow ($t \Omega = 70$).
The dots concentrated at $x/H \simeq 2.0$ in the figure are 
actual velocities of all the simulated particles.
The actual data of the particle velocities are well reproduced by 
equation~(\ref{eq:turb}), since most of the data points are located 
between the predicted minimum and maximum values.
We further decompose $v_{\rm d}$ into $v_{\rm f}$ and $v_{\rm t}$
according to equation~(\ref{eq:turb}).
Figure~\ref{fig:s40-dust}c presents the radial distribution 
of the two components. 
The maximum and minimum values for each component are plotted. 
It is shown that the effect of the remnant turbulence is 
small and overall features of particle radial migration is
represented by $v_{\rm f}$.
Due to the non-smooth pressure and velocity distributions,
${\rm min}[v_{\rm d}] \simeq {\rm min}[v_{\rm f}]>0$ at $-3.5 \lesssim x/H \lesssim -2.5$ and $x/H \sim -2.1$, in addition to $0.0 \lesssim x/H \lesssim 2.0$.
It means that all particles have positive radial velocity there and
accumulate near the outer edges of these regions.
On the other hand, ${\rm max}[v_{\rm d}] \simeq {\rm max}[v_{\rm f}]>0$ is observed in some areas, for example, around $x/H \simeq 3.0$. 
A small number of particles are stalled there. 

The peaks of particle density at $x/H<0$ would not grow because there are few particles which can accumulate from outside (Figure~\ref{fig:s40-dust}a). The stagnant areas cannot stall all the particles which migrate inward due to ${\rm min}[v_{\rm d}]<0$ (Figure~\ref{fig:s40-dust}b). Therefore the peak at $x/H \simeq 2.0$ is the most prominent concentration zone. 

\begin{figure}
\figurenum{9}
\epsscale{0.6}
\plotone{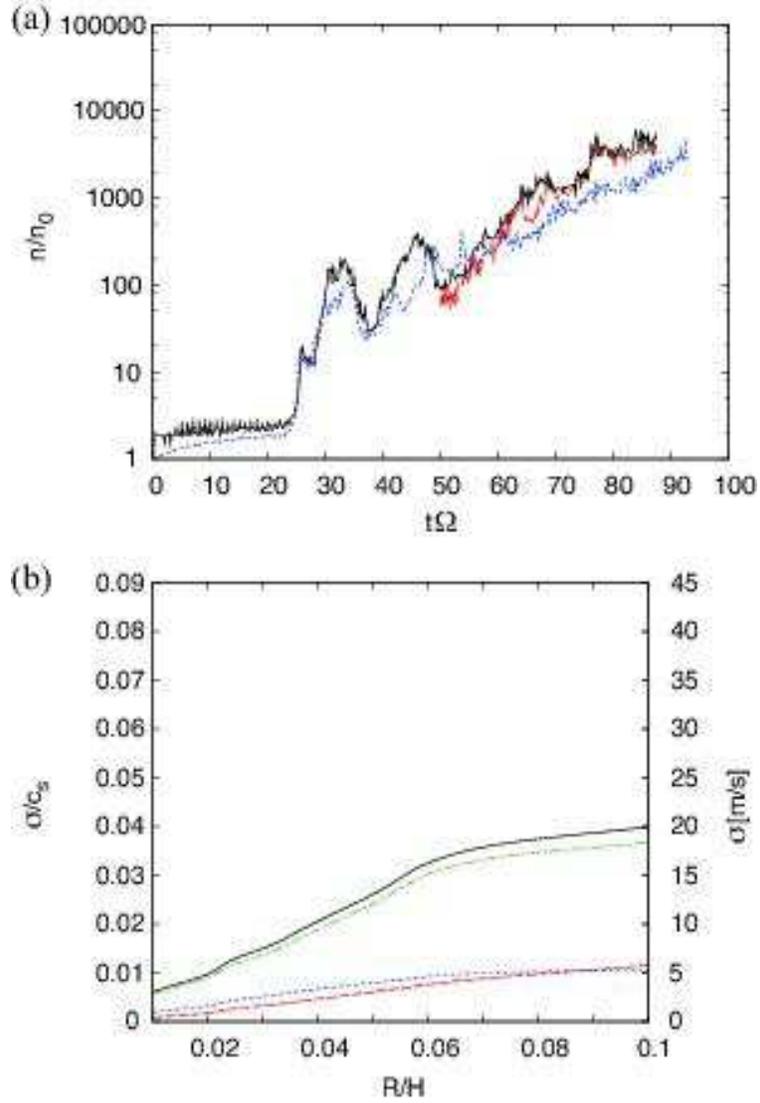}
\caption{The same plots as Figure~\ref{fig:eta-clump} but for model-s40 ($L_{\rm s}/H=4.0, \tau_f\Omega=1.0, \beta=-0.04$). The dotted line in Panel a is the number of particles in the densest cell in model-s40b ($\beta=-0.10$). }
\label{fig:s40-clump}
\end{figure}

Figure~\ref{fig:s40-clump}a shows the time variation of the number of particles in the densest grid in the whole region and at the center of the traced clump. The agreement between the two indicates that the same clump is growing. Figure~\ref{fig:s40-clump}b is the velocity dispersion of particles within the clump at $t=t_c (=55/\Omega)$. The velocity dispersion is small but is larger than in model-$\eta$. While the turbulence does not affect the location of the dust concentration, it does influence the degree of particle accumulation. 

When the azimuthal box size is long, shear velocity becomes supersonic near the edge and it may cause artificial density dip \citep{john08, joh09} unless a high-order scheme is applied for Keplerian advection term.
To confirm that the density dip found in our simulations is not caused by such a numerical error, we also carried out a run in which the unstable region is put in the side areas and the stable region is at the center.
The other parameters are the same as model-s40.
We found that the density dip is formed at the side area and no artificial density dip is created in the central area, which means that our numerical scheme using CIP method with the staggard mesh and the short timestep ($dt=10^{-4}$; Courant condition is set to be $<0.5$) is sufficiently high-order scheme.

\begin{figure}
\figurenum{10}
\epsscale{0.6}
\plotone{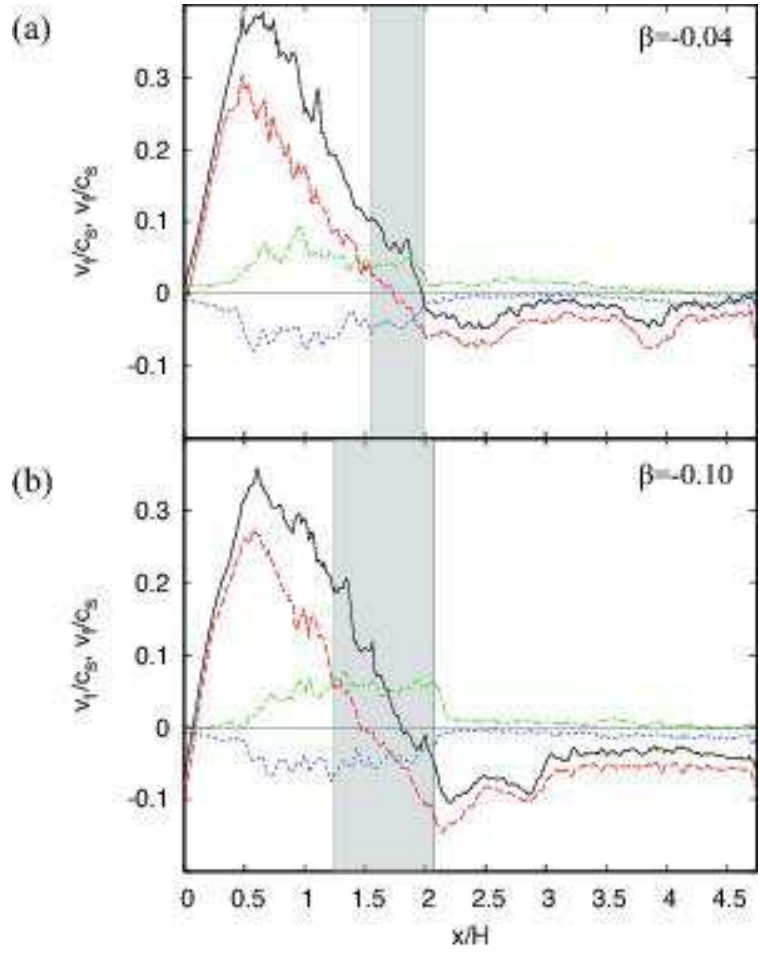}
\caption{(a) The closeup of Figure~\ref{fig:s40-dust}c (model-s40 in which $\beta=-0.04$). 
(b) The same plots with Panel a but for model-s40b in which $\beta=-0.10$.}
\label{fig:s40b-drift}
\end{figure}

Let us discuss the effects of a steeper global pressure gradient.  Following the same argument as in the final part of section 4.1, from the value of $v_{\rm f}$ or $\delta u_y$, a super-Keplerian region would remain unless the global prssure gradient is as anomalously large as $\beta<-0.64$.
To make a more quantitative argument, as was done in \citet{joh07b}, we also carried out a run with the pressure gradient of $\beta=-0.10$ (model-s40b). The other parameters are the same as model-s40. 
Figure~\ref{fig:s40b-drift} shows the close-up of the distribution of ${\rm max}, {\rm min}[v_{\rm f}]$ and ${\rm max}, {\rm min}[v_{\rm t}]$.
With the higher negative value of $\beta$, gas rotates more slowly and the stronger headwind deprives the dust particles of angular momentum more efficiently to accelerate their inward migration in the sub-Keplerian region. 
Since $v_{\rm t}$ does not depend on $\beta$, the value of $\left|v_{\rm f}\right|$ exceeds $\left|v_{\rm t}\right|$ by a larger degree in outer regions. Thus greater amounts of dust particles move to the dust concentration area without becoming locally stagnant due to non-smooth gas pressure and gas velocity fluctuations in the sub-Keplerian region. 
On the other hand, in the dust concentrated area, the dust particles are affected by turbulence more severely. With more downward shift of the $v_{\rm f}$ value in the whole region, the dust concentration that occurs at $v_{\rm f} \simeq0$ is shifted inward to $x/H\simeq1.8$ from $x/H\simeq2.0$. 
The magnetic turbulence is stronger at this new location and it increases the radial width ($\delta x$) of the dust accumulated region (denoted by ${\rm max}[v_{\rm d}]>0$ and ${\rm min}[v_{\rm d}]<0$: the gray hatched areas in the figure from $\delta x/H \simeq0.5$ to $\delta x/H \simeq 0.9$ and doubles the velocity dispersion of particles there. 
To see the overall consequences of these both positive and negative effects, the number of particles in the most dense grid cell is plotted in Figure~\ref{fig:s40-clump}a together with the result of model-s40. 
Despite the different dust radial velocity and gas turbulence, the density increase rate and the peak value are only slightly different from each other, although the new run could show higher dust density on longer timescale. 


\subsubsection{model-t01}

The effects of turbulence are larger for particles with shorter $\tau_f$. We have run a case with $\tau_f\Omega=0.1$ (model-t01). Except for the friction time, all the settings are the same as model-s40. Since there is no kick-back from the dust particles to the MHD fluid, MHD results are exactly the same (Figure~\ref{fig:s40-3D}a and Figure~\ref{fig:s40-puy}). The difference in the dust concentration pattern is caused by the enhanced gas drag. 

\begin{figure}
\figurenum{11}
\epsscale{0.6}
\plotone{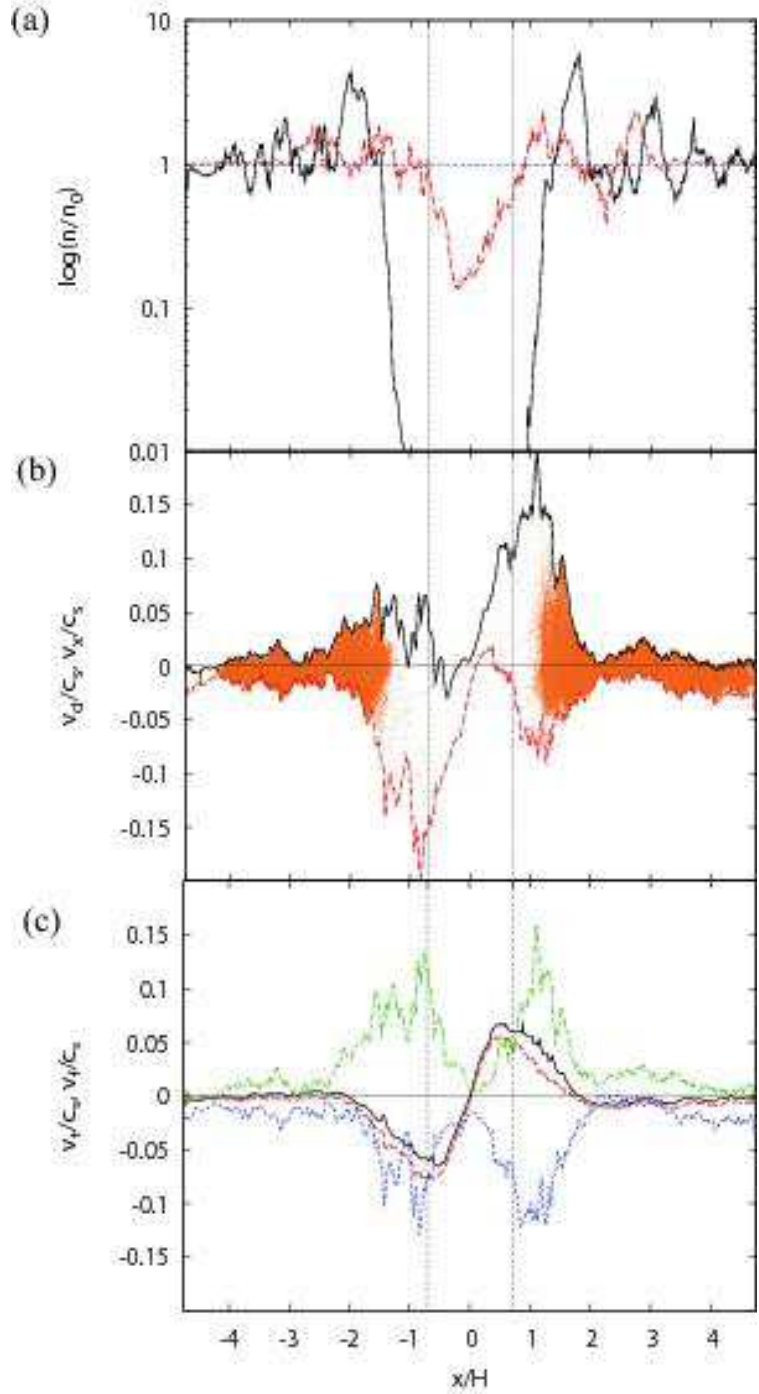}
\caption{The same plots as Figure~\ref{fig:s40-dust} but for model-t01 ($L_{\rm s}/H=4.0, \tau_f\Omega=0.1$), where turbulence effects are enhanced.}
\label{fig:t01-dust}
\end{figure}

Figure~\ref{fig:s40-3D}c presents the evolution of the particle density. The difference from the previous case that becomes visible at the later time is that the smaller dust particles are scattered over a broader area. Figure~\ref{fig:t01-dust}a shows the radial profile of the dust number density averaged azimuthally and vertically. Since the MHD flow patterns are exactly the same, the prominent density peaks are located at the same position as in model-s40. However, the peak values are lower and the radial widths are wider. Furthermore, more particles stay outside the peaks and are scattered in the stable region. 

\begin{figure}
\figurenum{12}
\epsscale{0.6}
\plotone{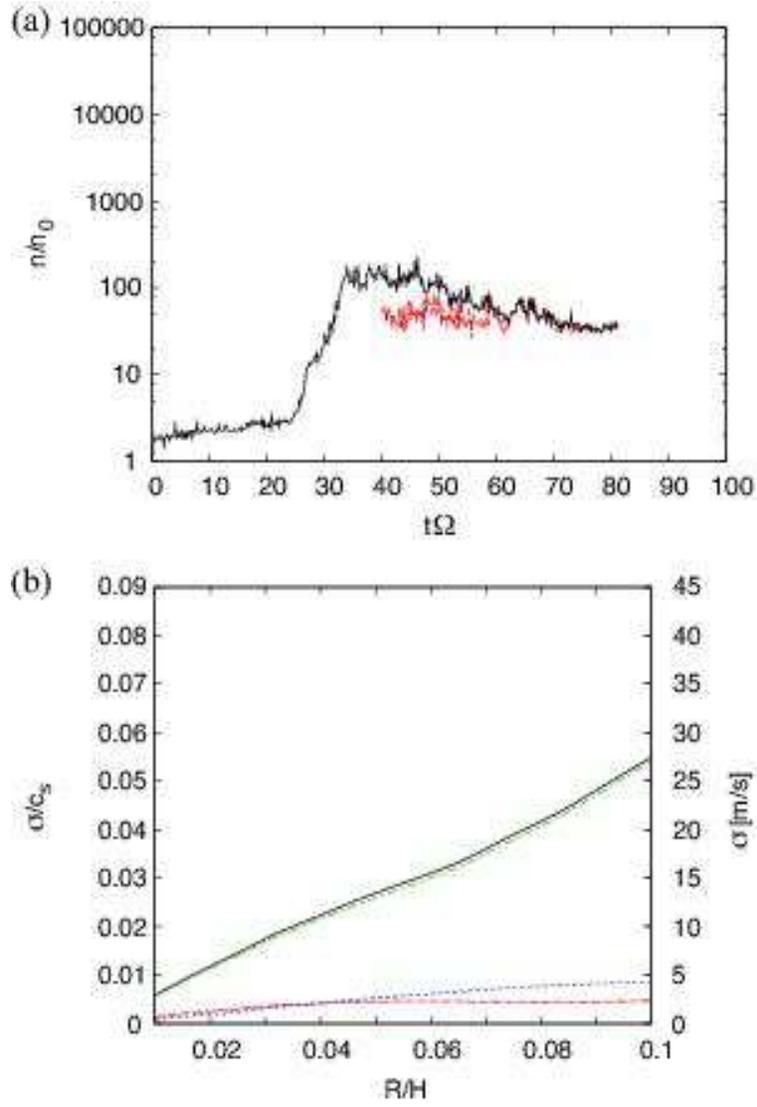}
\caption{The same plots as Figure~\ref{fig:eta-clump} but for model-t01 ($L_{\rm s}/H=4.0, \tau_f\Omega=0.1$). 
}
\label{fig:t01-clump}
\end{figure}

These differences are caused by the enhanced turbulent diffusion of dust particle motions through enhanced gas drag. Indeed, Figure~\ref{fig:t01-dust}c shows that the turbulent scattering $v_{\rm t}$ dominates over the steady migration $v_{\rm f}$ in most areas of the simulation box. 
Figure~\ref{fig:t01-dust}b shows the estimated radial velocity of particles $v_{\rm d}$. The radial velocity of the simulated particles, $v_{x}$, 
are also shown by dots. 
Although dust particles are swept out of the unstable region,
in most parts of the unstable region, 
max[$v_{\rm d}] > 0$ and min[$v_{\rm d}] < 0$ are satisfied, 
so the dust particles are broadly distributed in the stable region
with lower concentration.
Nevertheless, identity of the clump with the highest dust concentration is still maintained (Figure~\ref{fig:t01-clump}a). 
Indeed the velocity dispersion around center of the traced clump shown is not increased significantly (Figure~\ref{fig:t01-clump}b). 

We also run a high global pressure gradient model ($\beta=-0.10$). 
Though the inward migration of dust particles becomes faster on average, the maximum density is unchanged because the remnant turbulence that scatters the dust particles in the dust concentrated region is the dominant factor for this $\tau_f\Omega=0.1$ case.


\subsubsection{model-s055}

Recent works \citep[e.g.,][]{joh07a} proposed
that turbulence itself concentrates dust particles 
in turbulent eddies.
In their models, lifetime of the eddies has to be longer than
the timescale to form dense enough particle clumps for
formation of planetesimals.
On the other hand, our model proposes another path 
to accumulate dust particles that turbulence plays a role 
in transformation of the nearly Keplerian gas flow
into the quasi-steady flow with a local rigid rotation region.
Because the dust particles accumulate near the outer edge 
of super-Keplerian parts produced by the rigid rotation and 
the flow pattern is quasi-steady, 
the timescale problem does not exist in our model.
Actually, in the case of model-s40 and model-t01 
with $R_{\rm m,ave} \simeq 0.096$, 
the same clump with the highest density is maintained until
the end of simulations ($t \Omega \simeq 80$).

However, in our model, as $R_{\rm m,ave}$ increases to unity, 
stronger residual turbulence may destroy the clumps,
although the clumps are repeatedly created, 
which may inhibit planetesimal formation.
Here, we examine the cases with larger $R_{\rm m,ave}$ 
(but still $R_{\rm m,ave}<1$)
to find the critical value of $R_{\rm m,ave}$ for persistent clumping.

\begin{figure}
\figurenum{13}
\epsscale{1.0}
\plotone{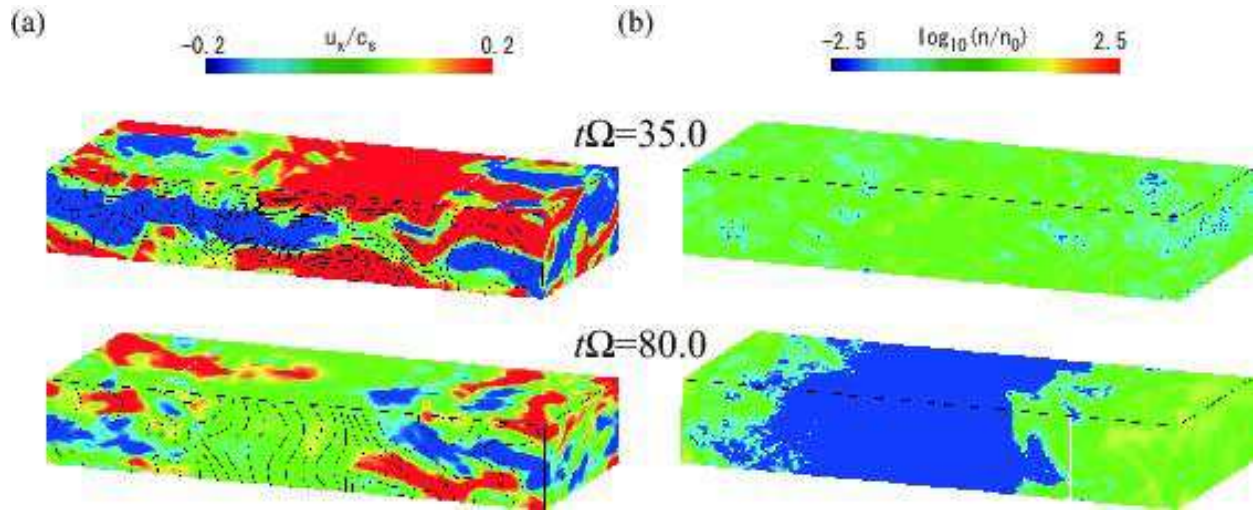}
\caption{The same plots as Figure~\ref{fig:eta-3D} but for model-s055 ($L_{\rm s}/H=0.55, \tau_f\Omega=1.0$). 
}
\label{fig:s055-3D}
\end{figure}

\begin{figure}
\figurenum{14}
\epsscale{1.0}
\plotone{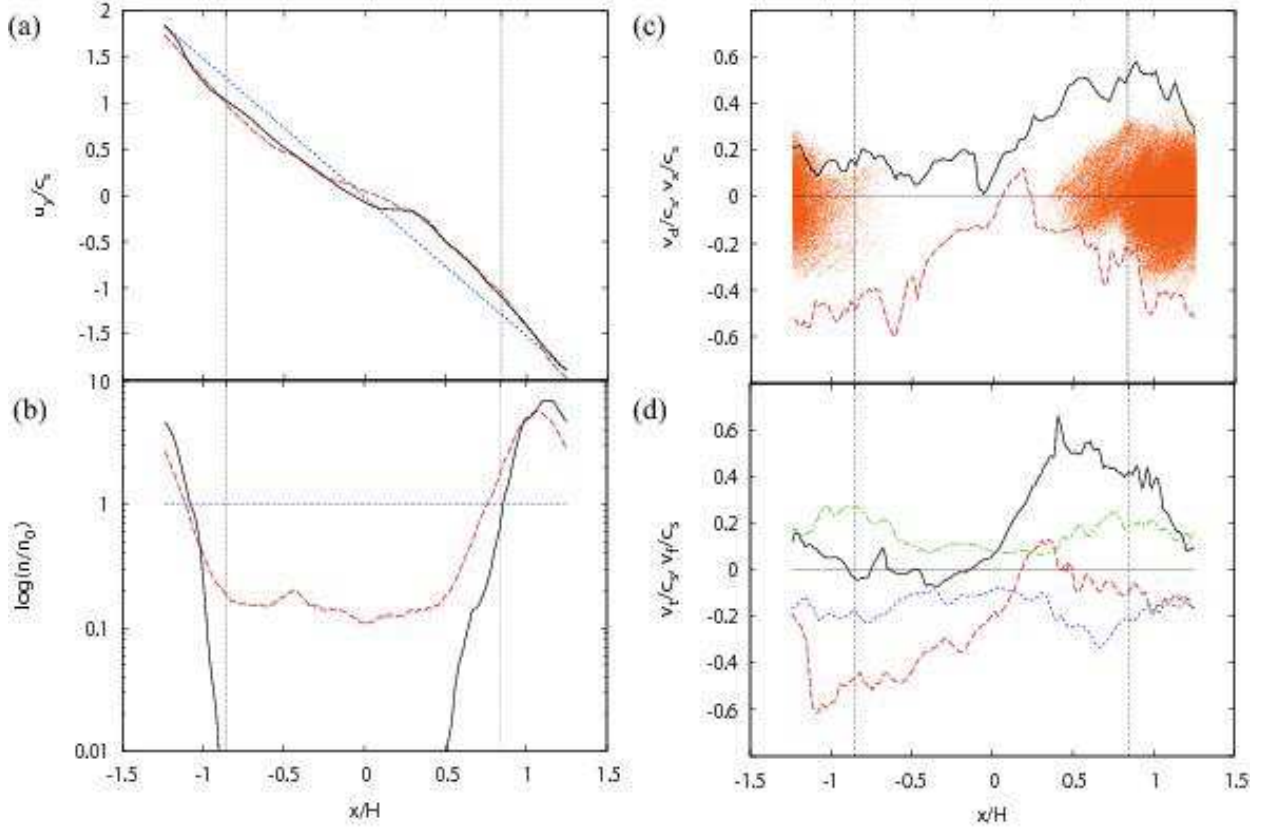}
\caption{(a) Time evolution of both azimuthally and vertically averaged angular velocity of gas $u_{y}$ in model-s055 ($L_{\rm s}/H=0.55, \tau_f\Omega=1.0$), where the dotted, dashed and bold lines represent the snapshots at $t\Omega= 0, 40$ and $70$, respectively. (b)(c)(d) The same plots as Figure~\ref{fig:s40-dust}a, b, and c but for model-s055. Dust particles are scattered due to elevated turbulence.}
\label{fig:s055-puy}
\end{figure}

In model-s055, the width of the stable region $L_{\rm s}$ is one eighth of model-s40.
Accordingly, $R_{\rm m,ave}=0.64$ compared to 0.096 of model-s40. 
Figure~\ref{fig:s055-3D}a shows that the MRI turbulence intrudes into the stable region and covers the whole region without being dissipated. By $t\Omega=80$, the instability is weakened.
Although a clear rigid rotation is not formed inside the unstable region (Figure~\ref{fig:s055-puy}a) because of the stronger turbulence 
that remains especially in the stable region, 
the velocity gradient becomes small inside the unstable region
and the super-Keplerian zone is formed.
As a result, the dust particles are swept out of the unstable region
and accumulate near the outer stable/unstable boundary (Figure~\ref{fig:s055-3D}b).

\begin{figure}
\figurenum{15}
\epsscale{0.6}
\plotone{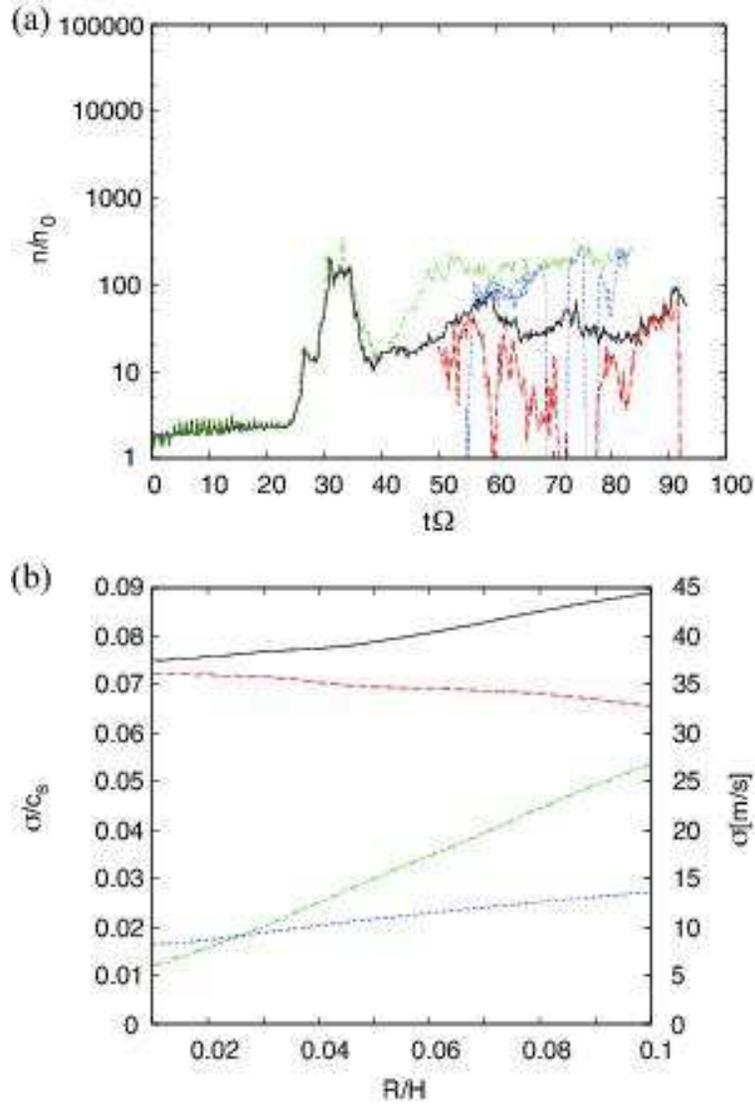}
\caption{The same plots as Figure~\ref{fig:eta-clump} but for model-s055 ($L_{\rm s}/H=0.55, \tau_f\Omega=1.0$). The thin dash-dotted and thin dotted lines in Panel a show the results of model-s011 ($L_{\rm s}/H=1.1, \tau_f\Omega=1.0$).
}
\label{fig:s055-clump}
\end{figure}

However, because the unstable region occupies larger fraction of the simulation box, the resultant magnetic field diffused out of the unstable region is larger than in the previous models. 
These lead to stronger remnant turbulence. 
Figure~\ref{fig:s055-clump}a shows that the number of particles in the densest grid of the whole region does not agree with that at the center of the traced clump. 
It means that the clumping is only tentative, although clumps are
repeatedly created.
The diffusive nature of the clumps in model-s055 is depicted by the larger velocity dispersion shown in Figure~\ref{fig:s055-clump}b. 

Thin lines in Figure~\ref{fig:s055-clump}a shows the clump identity test for model-s11 ($R_{\rm m,ave}=0.37$).
It shows that the clumps are not persistent as well.
Therefore, we conclude that
persistent clumps of dust particles are formed for $R_{\rm m,ave} \la 0.1$.

If the global pressure gradient is steeper, the radial velocity of dust particles is faster and its effect may seem to compete better with that of gas turbulent motion. 
However, from Figure~\ref{fig:s055-puy}d, the minimum value of $v_{\rm f}$ becomes negative in the entire regions if $\beta<-0.24$ and no particles can be accumulated at the boundary of super/sub-Keplerian regions. There is not much parameter space where more dust concentration can be expected.


\section{Conclusion and Discussion}\label{sec:discussion}

We have investigated the dust concentration in protoplanetary gas disks that are in the quasi-steady state created by inhomogeneous magnetorotational instability. We set the inhomogeneous instability with the initial radial non-uniformity
of either ionization degree or vertical component of the magnetic field. 
The result of our local three-dimensional resistive MHD simulations with Lagrangian particles clearly shows that the meter or decimeter sized dust particles are concentrated strongly and steadily in either setting
except for the cases with $R_{\rm m,ave} > 0.1$
(where $R_{\rm m,ave}$ is the averaged magnetic Reynolds number averaged over the simulation box), which can allow planetesimal formation via gravitational instability.

The MRI growth in our model generates the locally confined super-Keplerian region and this flow is quasi-stable due to the support of gas pressure gradient which is also formed by the MRI. 
Because unperturbed gas flow is sub-Keplerian due to global pressure gradient, the particles suffering gas drag are concentrated in the Keplerian domain at the outer-edge of the super-Keplerian area. 
Since the flow is in a quasi-steady state, particles are supplied from outer regions by gas drag migration and the particle density increases significantly.

The process of particle concentration and the increasing rate of dust density depend on the particle size, the initial radial widths of unstable/stable regions, and the initial non-uniformity setting as follows:
\begin{description}
\item[
\textit{The initial non-uniform setting, magnetic field or 
resistivity}:]
The non-uniform resistivity model produces more stable and clean state of local rigid rotation than the non-uniform magnetic field model, because the high resistivity dissipates the magnetic perturbations more rapidly in the stable regions.
Indeed, the velocity dispersion of particles in clumps is smaller in the non-uniform resistivity model. 
In the non-uniform magnetic field cases, the magnetic reconnection and the remnant weak instability create less smooth gas velocity field, so that particles tend to assemble at multiple radial positions. 
On the other hand, in the non-uniform resistivity case, all particles are concentrated at a specific radial position, and the peak values
of the particle density in clumps is higher, although the particle density also depends on the size of numerical box. 

\item[\textit{The dust size, meter or decimeter}:]
The meter-size particles are less coupled with gas motion than decimeter-size ones. Since their radial migration is faster, the meter-sized particles 
are more likely to be condensed in the radially narrow area without being scattered by the remnant gas turbulence.  
The peak dust density becomes $\ga 1000$ times larger than the initial value. 
Even in the decimeter-size particle cases, the velocity dispersion of particles in the clump excited by the remnant turbulence is sufficiently smaller than radial drift velocity due to the quasi-steady gas flow and dust concentration is still observed
in the simulations, although the enhanced dust density is at most $\sim 100$.

\item[\textit{The degree of remnant turbulence}:]
When the stable region is initially smaller in the non-uniform magnetic field model (CASE2), that is, $R_{\rm m,ave}$ is larger, 
stronger turbulence tends to remain. 
For $R_{\rm m,ave} \la 1$, however, the super-Keplerian area continues to 
exist and clumps with density $\sim 100$ times larger than initial values
are robustly created.
Furthermore, for $R_{\rm m,ave} \la 0.1$, the clumps are persistent
enough for following gravitational instability, while they are 
tentative and repeatedly produced in the cases of $0.1 \la R_{\rm m,ave} \la 1$. 
\end{description}

We should address the following points in subsequent papers:
\begin{description}
\item[\textit{consistent magnetic resistivity}:]
We need to calculate the value of magnetic resistivity consistently
as a time dependent value, because it is sensitive to both gas and dust densities.
Around strong dust concentration areas, gas and dust density become larger and magnetic resistivity becomes higher, which may calm down the remnant turbulence.

\item[\textit{Rossby wave instability (RWI)}:] 
RWI can be caused from a pressure bump on the radial-azimuthal plane. If we simulate in an enlarged box, vortex may grow, although in the non-uniform magnetic field cases, the growth of the instability may be reduced by the azimuthal magnetic field.
If RWI occurs, it may serve as a mechanism of dust concentration 
in the azimuthal direction \citep{ina06,lyra08}.

\item[\textit{the vertical structure of a protoplanetary disk}:]
In the upper disk regions, ionization degree may be so high that the MRI 
to occur \citep{fle03, fro06, oishi07}. 
The turbulence there may affect the small dust dynamics 
near the midplane. 

\item[\textit{feedback from dust particles onto gas}:]
The feedback may be the most important.
A great concentration in our results here will inevitably change gas velocity field and it could affect the condition of dust concentration (density enhancement, velocity dispersion and so on) and planetesimal formation.
We have already started simulations with this effect.

\item[\textit{collisional destruction}:]
While the velocity dispersions of dust particles in clumps are extremely low in some of our results, the dust particles can be collisionally disrupted into small fragments that are coupled to gas motion and the fragments may be dispersed by even weak turbulence. 
For clumps to gravitationally collapse, its timescale needs to be smaller than mean collision time.  
On the other hand, if the dust grains are also produced by the collisions, they might affect the gas ionization state through recombination of electrons onto their surface.
We need to address these feedback effects by collisional disruption as well.

\end{description}

\acknowledgments
We thank the referees for insightful and helpful comments. This work was supported by Grant-in-Aid for JSPS Fellows (208778). The simulations presented in this paper were performed by NEC SX-6 at ISAS/JAXA.


\clearpage

\clearpage

\clearpage

\clearpage

\clearpage

\clearpage

\clearpage

\clearpage

\clearpage

\clearpage

\clearpage

\clearpage

\clearpage

\clearpage

\clearpage

\begin{table}
\begin{tabular}{lccclllll}
\hline \hline
model-& $L_{x}\times L_{y}\times L_{z}\left(/H\right)$ &
$N_{x}\times N_{y}\times N_{z}$ & 
\parbox[c]{6em}{non-uniform parameter} & 
$L_{\rm u}/H$ & $L_{\rm s}$/H & $R_{\rm m,ave}$ & 
$\tau_{f}\Omega$ & $N_{\rm d}/10^{7}$ \\ \hline

$\eta$ & $1.4\times1.0\times0.14$ & $280\times200\times28$
             & $\eta$ & 0.28 & 0.56 & *** & 1.0 & 1.3 \\
s40 & $9.5\times1.0\times0.50$ & $950\times100\times50$
             & $\mathbf{B}$ & 1.4 & 4.0 & 0.096 & 1.0 & 3.8 \\
s11 & $3.5\times1.0\times0.50$ & $350\times100\times50$
             & $\mathbf{B}$ & 1.4 & 1.1 & 0.37 & 1.0 & 1.4 \\
s055 & $2.5\times1.0\times0.50$ & $250\times100\times50$
             & $\mathbf{B}$ & 1.4 & 0.55 & 0.64 & 1.0 & 1.0 \\
t01 & $9.5\times1.0\times0.50$ & $950\times100\times50$
             & $\mathbf{B}$ & 1.4 & 4.0 & 0.096 & 0.1 & 3.8 \\
\hline
\end{tabular}

\caption{RUN PARAMETERS
(1): Name of run. (2): Size of simulation box. 
(3): Grid resolution. (4): Parameter set non-uniformly.
(5): Radial width of unstable region. 
(6): Radial width of stable region. 
(7): Friction time. (8): Number of particles.
}
\label{tab:1}
\end{table}

\clearpage

\begin{table}
\begin{tabular}{lllrll}
\hline \hline
model- & $t_{c}\Omega$ & $\Omega\delta t_{trace}$ & $N_{\rm c}$ &
$\sigma_{\rm p,ave}\left(\sigma_{{\rm p},x}, \sigma_{{\rm p},y}, 
\sigma_{{\rm p},z}\right)/H$ \\ \hline

$\eta$ & 55.0 & 34.0 & 71096 & 0.013(0.0023, 0.012, 0.0036) \\
s40 & 75.0 & 37.4 & 13823 & 0.16(0.041, 0.14, 0.054) \\
s11 & 75.0 & 33.6 & 2285 & 0.77(0.67, 0.24, 0.13) \\
s055 & 90.0 & 43.2 & 392 & 0.65(0.57, 0.24, 0.13) \\
t01 & 75.0 & 41.2 & 292 & 0.18(0.086, 0.13, 0.069) \\
\hline
\end{tabular}

\caption{TRACED PARTICLES
(1): Name of run. (2):Time to define which particles are traced. 
(3): Period for tracing. (4): Number of traced particles. 
(5): Position dispersion of $N_{\rm c}$ particles 
averaged for $\delta t_{trace}\Omega$ and its composition. 
}
\label{tab:2}
\end{table}

\end{document}